\documentclass[12pt]{article}
\usepackage[external,final,fonts]{cgpg}
\usepackage{graphicx}
\usepackage{psfrag}

\newcommand{\emA}{\mbb{A}}
\newcommand{\emF}{\mbb{F}}
\newcommand{\emE}{\mbb{E}}
\newcommand{\emB}{\mbb{B}}
\newcommand{\emT}{\mbb{T}}

\newcommand{\eme}{\mbb{e}}
\newcommand{\scri}{\mfs{I}}
\newcommand{\spi}{i^o}
\newcommand{\man}{\mathcal{M}}
\newcommand{\area}{\ensuremath a_\Delta^{}}
\newcommand{\rad}{\ensuremath r_\Delta^{}}
\newcommand{\ERad}{\ensuremath E_\infty^\mathrm{rad}}

\newcommand{\ADM}{{\mathrm{ADM}}}
\newcommand{\Rad}{{\mathrm{rad}}}

\newcommand{\dual}{{}^\star}

\def\U{{\rm U}}
\def\SU{{\rm SU}}

\def\SL{{\rm SL}}

\def\IH{\mathcal{IH}}
\def\S{\mathcal{S}}

\newcommand{\half}[0]{\ensuremath \tsfrac{1}{2}}

\renewcommand{\i}[0]{\ensuremath \iota}
\renewcommand{\o}[0]{\ensuremath o}

\begin{document}

\title  {Mechanics of Isolated Horizons}
\author {Abhay Ashtekar${}^{1,2}$, Christopher Beetle${}^{1}$
  and Stephen Fairhurst${}^{1}$\\
\address{1. Center for Gravitational Physics and Geometry\\
            Department of Physics, The Pennsylvania State University\\
            University Park, PA 16802\\
         2. Institute for Theoretical Physics,\\
            University of California, Santa Barbara, CA 93106, USA}}
\date   {November 4, 1999}
\maketitle

\begin{abstract}
A set of boundary conditions defining an \textit{undistorted,
non-rotating isolated horizon} are specified in general relativity.  A
space-time representing a black hole which is itself in equilibrium
but whose exterior contains radiation admits such a horizon.  However,
the definition is applicable in a more general context, such as
cosmological horizons.  Physically motivated, (quasi-)local
definitions of the mass and surface gravity of an isolated horizon are
introduced and their properties analyzed.  Although their definitions
do not refer to infinity, these quantities assume their standard
values in the static black hole solutions.  Finally, using these
definitions, the zeroth and first laws of black hole mechanics are
established for isolated horizons.
\end{abstract}

\section{Introduction}
\label{s1}

The similarity between the laws of black hole mechanics and those of ordinary
thermodynamics is one of the most remarkable results to emerge from classical
general relativity \cite{1,2,bc,3}.  However, the original framework was
somewhat restricted and it is of considerable interest to extend it in certain
directions, motivated by physical considerations.  The purpose of this paper
is to present one such extension.

The zeroth and first laws refer to equilibrium situations and small departures
therefrom.  Therefore, in this context, it is natural to focus on isolated
black holes.  In the standard treatments, these are generally represented by
\textit{stationary} solutions of field equations, i.e, solutions which admit a
time-translational Killing vector field \textit{everywhere}, not just in a
small neighborhood of the black hole.  While this simple idealization is a
natural starting point, it seems to be overly restrictive.  Physically, it
should be sufficient to impose boundary conditions at the horizon which ensure
\textit{only that the black hole itself is isolated}.  That is, it should
suffice to demand only that the intrinsic geometry of the horizon be time
independent, whereas the geometry outside may be dynamical and admit
gravitational and other radiation.  Indeed, we adopt a similar viewpoint in
ordinary thermodynamics; in the standard description of equilibrium
configurations of systems such as a classical gas, one usually assumes that
only the system under consideration is in equilibrium and stationary, not the
whole world.  For black holes in realistic situations, one is typically
interested in the final stages of collapse where the black hole is formed and
has `settled down' or in situations in which an already formed black hole is
isolated for the duration of the experiment (see figure 1).  In such
situations, there is likely to be gravitational radiation and non-stationary
matter far away from the black hole, whence the space-time as a whole is not
expected to be stationary.  Surely, black hole mechanics should incorporate such situations.

\begin{figure}
  \begin{center}
    \begin{minipage}{2.4in}
      \begin{center}
        \psfrag{Ip}{$\scri^+$}
        \psfrag{calM}{$\mathcal{M}$}
        \psfrag{ip}{$i^+$}
        \psfrag{i0}{$\spi$}
        \psfrag{Delta}{$\Delta$}
        \psfrag{M}{$M$}
        \resizebox{!}{6cm}{\includegraphics{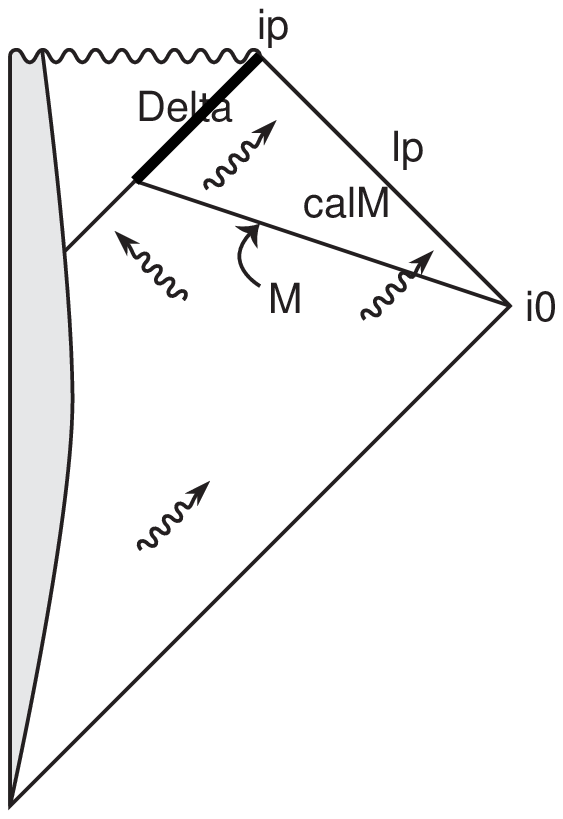}}\\(a)
      \end{center}
    \end{minipage}
    \hspace{.4in}
    \begin{minipage}{3.4in}
      \begin{center}\small
        \psfrag{D1}{$\Delta_2$}
        \psfrag{D2}{$\Delta_1$}
        \psfrag{i0}{$\spi$}
        \psfrag{H}{$\mfs{H}$}
        \includegraphics[width=8cm]{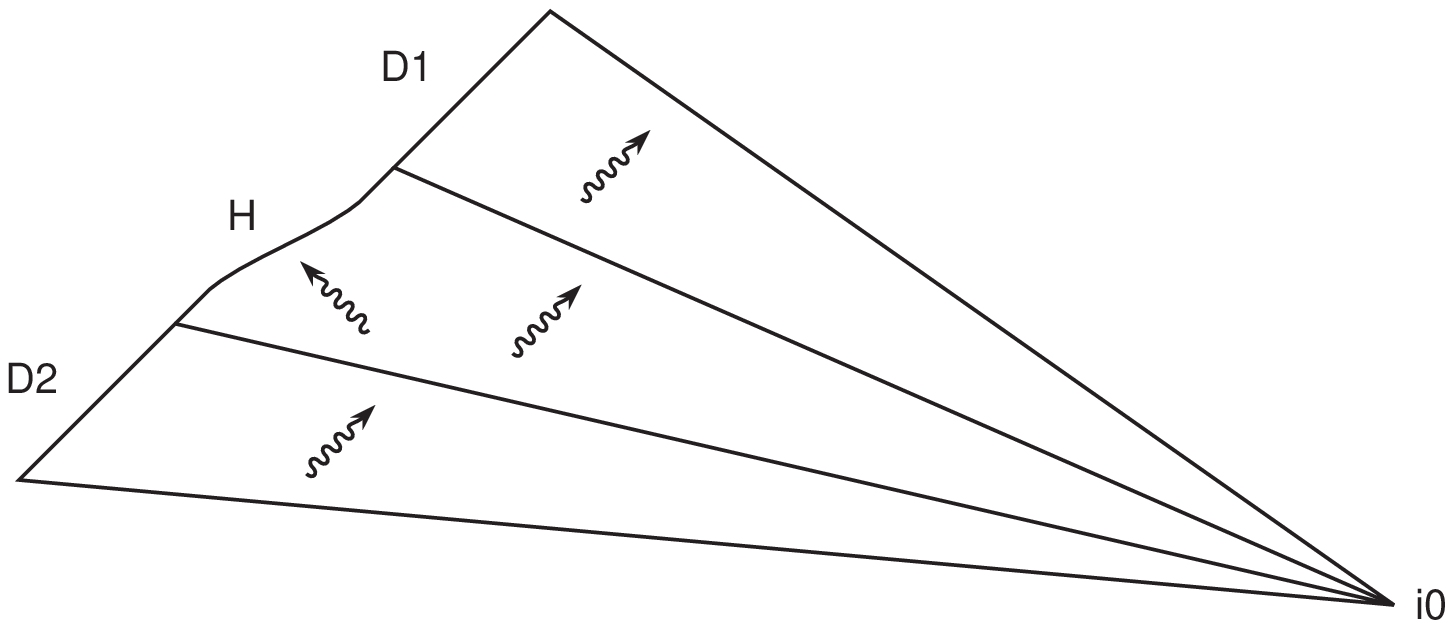}\\(b)
      \end{center}
    \end{minipage}
    \caption{(a)\quad A typical gravitational collapse.
The portion $\Delta$ of the horizon at late times is isolated.  The
space-time $\mathcal{M}$ of interest is the triangular region bounded
by $\Delta$, $\scri^+$ and a partial Cauchy slice $M$.  \quad(b)\quad
Space-time diagram of a black hole which is initially in equilibrium,
absorbs a small amount of radiation, and again settles down to
equilibrium.  Portions $\Delta_1$ and $\Delta_2$ of the horizon are
isolated.}\label{exam} \end{center}
\end{figure}

A second limitation of the standard framework lies in its dependence on
\textit{event} horizons which can only be constructed retroactively, after
knowing the \textit{complete} evolution of space-time.  Consider for example,
figure 2 in which a spherical star of mass $M$ undergoes a gravitational
collapse.  The singularity is hidden inside the null surface $\Delta_1$ at
$r=2M$ which is foliated by a family of marginally trapped surfaces and would
be a part of the event horizon if nothing further happens.  Suppose instead,
after a very long time, a thin spherical shell of mass $\delta M$ collapses.
Then $\Delta_1$ would not be a part of the event horizon which would actually
lie slightly outside $\Delta_1$ and coincide with the surface $r= 2(M+\delta
M)$ in the distant future.  On physical grounds, it seems unreasonable to exclude
$\Delta_1$ a priori from thermodynamical considerations.  Surely one should be
able to establish the standard laws of mechanics not only for the
event horizon but also for $\Delta_1$.

Another example is provided by cosmological horizons in de
Sitter space-time \cite{gh}.  In this case, there are no singularities or
black-hole event horizons.  On the other hand, semi-classical
considerations enable one to assign entropy and temperature to these
horizons as well.  This suggests the notion of event horizons is too
restrictive for thermodynamical analogies.  We will see that this is indeed
the case; as far as equilibrium properties are concerned, the notion of
event horizons can be replaced by a more general, quasi-local notion of
`isolated horizons' for which the familiar laws continue to hold.  The
surface $\Delta_1$ in figure 2 as well as the cosmological horizons in de
Sitter space-times are examples of isolated horizons.

\begin{figure}
  \begin{center}
    \includegraphics[height=4cm]{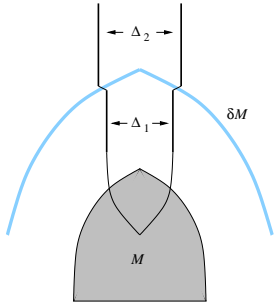}
\caption{A spherical star of mass $M$ undergoes collapse.  Later, a
spherical shell of mass $\delta{M}$ falls into the resulting black
hole.  While $\Delta_1$ and $\Delta_2$ are both isolated horizons,
only $\Delta_2$ is part of the event horizon.}\label{shell}
\end{center}
\end{figure}

In addition to overcoming these two limitations, the framework presented here
provides a natural point of departure for quantization and entropy
calculations \cite{6,ack,abk}.  In contrast, standard treatments of black hole
mechanics are often based on differential geometric identities and are not
well-suited to quantization.

At first sight, it may appear that only a small extension of the standard
framework based on stationary event horizons is needed to overcome the
limitations discussed above.  However, this is not the case.  For example, in
the stationary context, one identifies the black-hole mass with the ADM mass
defined at spatial infinity.  In the presence of radiation, this simple
strategy is no longer viable since even radiation fields well outside the
horizon also contribute to the ADM mass.  Hence, to formulate the first law, a
new definition of the black hole mass is needed.  Similarly, in the absence of
a globally defined Killing field, we need to generalize the notion of surface
gravity in a non-trivial fashion.  Indeed, even if space-time happens to be
static in a neighborhood of the horizon --- already a stronger condition than
contemplated above --- the notion of surface gravity is ambiguous because the
standard expression fails to be invariant under constant rescalings of the
Killing field.  When a \textit{global} Killing field exists, the ambiguity is
removed by requiring the Killing field be unit at \textit{infinity}.  Thus,
contrary to intuitive expectation, the standard notion of the surface gravity
of a stationary black hole refers not just to the structure at the horizon,
but also to infinity.  This `normalization problem' in the definition of the
surface gravity seems especially difficult in the case of cosmological
horizons in (Lorentzian) space-times whose Cauchy surfaces are compact.  Apart
from these conceptual problems, a host of technical issues must also be
resolved.  In Einstein--Maxwell theory, the space of stationary black hole
solutions is finite-dimensional whereas the space of solutions admitting
isolated horizons is \textit{infinite}-dimensional since these solutions also
admit radiation.  As a result, the introduction of a well-defined action
principle is subtle and the Hamiltonian framework acquires qualitatively new
features.

The organization of this paper is as follows.  In Section \ref{s2} we recall
the formulation of general relativity in terms of $\SL(2,\Com)$-spin soldering
forms and self-dual connections for asymptotically flat space-times
\textit{without} internal boundaries.  In Section \ref{s3}, we specify the
boundary conditions which define non-rotating isolated horizons and discuss
several examples.  The primary focus in Section 3, as in the rest of the
paper, is on Einstein--Maxwell theory, although more general matter is also
considered.  The consequences of these boundary conditions which are needed to
obtain the laws governing isolated horizons are discussed in Section \ref{s4}. 
Using this structure, we introduce in Section \ref{s5} the notion of the
surface gravity $\kappa$ of an isolated horizon without any reference to
infinity and prove the zeroth law.

The action principle and the Hamiltonian formulation are discussed in Section
\ref{s6}.  The Hamiltonian has a bulk term and two surface terms, one at
infinity and the other at the isolated horizon $\Delta$.  The bulk term is the
standard linear combination of constraints and the surface term at infinity
yields the ADM energy.  In Section \ref{s7}, we argue the horizon surface term
in the Hamiltonian should be identified with the mass $M_\Delta$ of the
isolated horizon.  In particular, in the situation depicted in figure 1(a) we
show $M_\Delta$ is the difference between the ADM energy and the energy
radiated away through all of null infinity (provided certain assumptions on
the structure at $i^+$ hold).  That is, although the definition of $M_\Delta$
uses only the structure available at the horizon, it equals the future limit
of the Bondi mass defined entirely on $\scri^+$ in the case under
consideration.
This is precisely what one might expect on physical grounds in the presence of
radiation.  Finally, having $\kappa$ and $M_\Delta$ at our disposal, we
establish the first law in Section \ref{s8}.

The overall viewpoint and the boundary conditions of Section \ref{s3}
are closely related to those introduced by Hayward in an interesting
series of papers \cite{9} which also aims at providing a more physical
framework for discussing black holes.  There is also an inevitable
overlap between this paper and \cite{ack} in which a Hamiltonian
framework (in terms of \textit{real} $\SU(2)$ connections) is
constructed with an eye towards quantization.  However, by and large
the results presented here are complementary to those obtained in
\cite{ack}.  Even when there is an overlap, the material is presented
from a different angle.  The relation between these three papers is
discussed in Section \ref{s9}.  Technical, background material needed
at various stages is collected in the three appendices.

A brief summary of our main results was given in \cite{abf1}.

\section{Mathematical Preliminaries}
\label{s2}

In this paper we will use the formulation of general relativity as a
dynamical theory of connections \cite{16} rather than metrics.
Classically, the theory is unchanged by the shift to connection variables.
However, at present the connection variables appear to be indispensable for
non-perturbative quantization \cite{blue}.  In particular, the entropy
calculation for isolated horizons has been carried out only within this
framework \cite{abk}.  Therefore, for uniformity, we will use the
connection variables in the main body of this paper.  However, as indicated
in Section \ref{s9}, all these results can also be derived in the framework
of tetrad dynamics.

In order to fix notation and conventions and to acquaint the reader with
the basics of connection dynamics, we will provide here a brief review of
this formulation in an asymptotically flat space-time without interior
boundary.  (The modifications required to accommodate a non-zero
cosmological constant are discussed at the end.)  For further details see,
e.g., \cite{blue}.

\subsection{Connection Variables}
\label{s2.1}

Fix a four-dimensional manifold $\man$.  In Einstein--Maxwell theory, the basic
fields consist of the triplet of asymptotically flat, smooth fields
$(\tensor<\sigma_a^AA'>, \tensor<A_a^AB>, \emA_a)$.  Here,
$\tensor<\sigma_a^AA'>$ is an anti-Hermitian soldering form for $\SL(2,\Com)$
spinors, $\tensor<A_a^AB>$ is a self dual $\SL(2,\Com)$ connection acting only on
unprimed spinors and $\emA_a$ is the $\U(1)$ electro-magnetic connection.  The
action for the theory is given by
\begin{equation}\label{afAct}
\tilde{S}(\sigma, A, \emA)
=\frac{-i}{8\pi G} \int_\man \Tr [\Sigma \wedge
F]+\frac{1}{8 \pi} \int_\man \emF \wedge \dual \emF +\frac{i}{8
\pi G} \int_{C_\infty} \Tr [\Sigma \wedge A].
\end{equation}
The 2-form fields $\Sigma$ are given by $\Sigma^{AB}=\sigma^{AA'} \wedge
\tensor<\sigma_A'^B>$, $F$ is the curvature of the connection $A$, $\emF$ is
the electro-magnetic field strength, and $C_\infty$ is the time-like cylinder
at spatial infinity.  We define a metric $g_{ab}$ of signature $-+++$ on this
manifold via $g_{ab} =\tensor<\sigma_a^AA'> <\sigma_bAA'>$.  With respect to
this metric, the 2-form fields $\Sigma$ are self dual.  (For details, see
\cite{blue,ar}.)

Let us consider the equations of motion arising from this action.
Varying the action with respect to the connection $A$, one obtains
\begin{equation}\label{stGauss}
D\Sigma=0.
\end{equation}
This implies the connection $D$ defined by $A$ has the same action on
unprimed spinors as the self dual part of the connection $\nabla$
compatible with the soldering form, $\nabla_a
\tensor<\sigma_b^AA'>=0$.  When this equation is satisfied, the
curvature $F$ is related to the Riemann curvature by:
\begin{equation} \label{1}
  \tensor <F_ab^AB> `=  - \frac{1}{4}' <\Sigma_cd^AB> <R_ab^cd>.
\end{equation}
Varying the action (\ref{afAct}) with respect to $\sigma$ and taking
into account the compatibility of $A$ with $\sigma$, we obtain a
second equation of motion
\begin{equation}\label{eom2}
G_{ab}= 8\pi G \emT_{ab}
\end{equation}
where $G_{ab}$ is the Einstein tensor of $g_{ab}$ and $\emT_{ab}$ is the
standard stress-energy tensor of the Maxwell field $\emF$
\cite{art,blue}.

Next, let us consider the equations of motion for the Maxwell field,
\begin{equation} \label{maxeq}
 d\emF =0, \quad\quad {\rm and} \quad\quad d\dual\emF= 0.
 \end{equation}
Since $\emF$ is the curvature of the $\U(1)$ connection $\emA$, the first
Maxwell equation $d\emF =0$ is an identity.  If one varies equation
(\ref{afAct}) with respect to $\emA$, one obtains the second Maxwell equation,
$d \dual \emF = 0$.  Thus, the equations of motion which follow from the
action (\ref{afAct}) are the same as those given by the usual
Einstein--Hilbert--Maxwell action; the two classical theories are equivalent.

\subsection{Hamiltonian Formulation}
\label{s2.2}

To pass to the Hamiltonian description of the theory, it is necessary to
re-express the action in terms of 3-dimensional fields.  Let us assume the
space-time $\mathcal{M}$ is topologically $M\times \Re$.  Introduce a `time
function' $t$ which agrees with a standard time coordinate defined by the
asymptotically Minkowskian metric at infinity.  A typical constant $t$ leaf of
the foliation will be denoted $M$.  Fix a smooth time-like vector field $t^a$,
transverse to the leaves $M$ such that: (i) $t^a \nabla_a t=1$ and (ii) $t^a$
tends to the unit time translation at spatial infinity.  We will denote the
future directed, unit normal to the leaves $M$ by $\tau^a$.  The intrinsic
metric on the 3-surfaces $M$ is $g_{ab} := \tensor ^4<g_ab>
+\tau_a \tau_b$.%
\footnote{ In the discussion of the Legendre transform and the Hamiltonian,
both in this section and section \ref{s6}, the four dimensional fields will
carry a superscript $^4$ preceding the field and all other fields will be
assumed to be three-dimensional, living on the space-like surface $M$.}
As usual, by projecting $t^a$ into and orthogonal to $M$, we obtain the the
lapse and shift fields, $N$ and $N^a$ respectively: $t^a=N\tau^a + N^a$.

We are now in a position to define the basic phase space variables of the
theory.  They are simply the pull backs to the space-like 3-surfaces $M$ of
the space-time variables $A, \Sigma$ and $\emA$, together with the electric
field two-form $\emE$.  To perform the Legendre transform, note the pull back
of the soldering form $\sigma$ to $M$ induces an $\SU(2)$ soldering form and
the pull back of the connection $A$ induces a complex-valued, $\SU(2)$
connection on spatial spinors.  More precisely, we have
\begin{equation}
\begin{eqtableau}{3}\label{3+1def}
\tensor<\sigma_a^{AB}> &= -i\sqrt{2}\,\tensor<g^b_a> ^4<\sigma_b^{AA'}>
<\tau_A'^B>
&&\qquad\Longleftrightarrow\qquad
&\tensor^4<\sigma_a^AA'> &=i\sqrt{2}\,
\tensor<\sigma_a^A_B> <\tau^BA'> - \tensor<\tau_a> <\tau^AA'> \\
\tensor^4<A_A^B> &=\tensor<A_0A^B>dt + \tensor<A_A^B>
&&\hfil\mbox{and}
&\tensor^4<\emA> &= \emA_0 dt +\emA,
\end{eqtableau}
\end{equation}
where $\tensor <\tau^AA'> `:=' ^4<\sigma_a^AA'> \tau^a$, $\tau^a \tensor
^4<A_a^AB> = 0$ and $\tau^a \tensor ^4<\emA_a> = 0$.  (Note that this
decomposition of connection 1-forms uses the space-time metric.)  Using these
definitions, one arrives at the following 3+1 decompositions of $\Sigma$ and
the gravitational and electro-magnetic field strengths:
\begin{equation}
\begin{eqtableau}{1}\label{3+1field}
\tensor^4<\Sigma> &= \Sigma -iN\sqrt{2}\,\sigma\wedge dt\\
^4F &= (-\dot{A} + D(t.^4A) +\vec{N}\vins F)\wedge dt + F \\
^4\emF &= (-\dot{\emA} +d(t.^4\emA)+\vec{N}\vins d\emA)\wedge dt +
d\emA,
\end{eqtableau}
\end{equation}
where $\vec{N}$ is the shift field.  The electric field $\emE$ and
magnetic field $\emB$ are defined as usual to be the pull backs to the
space-like hypersurface $M$ of $\dual\emF$ and $\emF$ respectively.

\let\puto=\overcirc

The phase space of the theory consists of quadruples $(\tensor<A_a^AB>,
\tensor<\Sigma_ab^AB>,\emA_a,\emE_{ab})$.  These fields satisfy the standard
falloff conditions \cite{blue}.  To specify them, let us fix an $\SU(2)$
soldering form $\puto{\sigma}$ on $M$ such that the 3-metric $\puto{g}_{ab} =
\puto{\sigma}_{a}{}^{AB} \puto{\sigma}_{bAB}$ is flat outside of a compact
set.  Then the quadruple of fields is required to satisfy:
\begin{equation}\label{afbc}
  \begin{eqtableau}{1}
    \Sigma_{ab} - \left( 1 + {M(\theta, \phi)\over r} \right) \puto{\Sigma}_{ab}
      &= O\left({1\over r^2}\right),\\
    A_a + {1\over 3} \Tr \left[ \puto{\sigma}^m A_m \right] \puto{\sigma}_a 
      &= O\left({1\over r^2}\right),\\
    \Tr \left[ \puto{\sigma}^a A_a \right] &= O\left({1\over r^3}\right),\\
  \end{eqtableau}
  \qquad
  \begin{eqtableau}{1}
    \emA &= O \left( \frac{1}{r^2} \right),\\
    \emE &= O \left( \frac{1}{r^2} \right).\\
\end{eqtableau}
\end{equation}
where $r$ is the radial coordinate defined by the flat metric $\puto{g}_{ab}$.

The action can be re-expressed in terms of these fields using equations
(\ref{3+1def}) and (\ref{3+1field}).  {}From this action, it is
straightforward to read off the Hamiltonian and symplectic structure.
\begin{equation}\label{AFPhSp}
   \begin{eqtableau}{1}
   \tilde{H}_t &= \int_M \frac{-i}{8\pi G}\Tr [(t.^4A)D\Sigma]
      +\frac{1}{4\pi}(t.^4\emA)d\emE
      +\left[ \frac{i}{8\pi G}\Sigma \wedge(\vec{N}\vins F)
      -\frac{1}{4\pi} \emE \wedge(\vec{N} \vins d\emA) \right] \\
    &\qquad\qquad +\frac{N}{8\pi G}(\Tr[\sqrt{2} \sigma \wedge F]
        - G(\emE \wedge \dual\emE + d\emA \wedge \dual
d\emA)) \\
    &\qquad +\frac{1}{8\pi G} \oint_{S_{\infty}}
      Tr[\sqrt{2} N \sigma \wedge A + i(\vec{N}\vins A)\Sigma] \\
  \tilde\Omega(\delta_1, \delta_2) &= \frac{-i}{8\pi G} \int_M
    \Tr[\delta_1 A \wedge \delta _2 \Sigma-
    \delta_2 A \wedge \delta _1 \Sigma]
    + \frac{1}{4\pi} \int_{M}\delta_1 \emA \wedge \delta_2 \emE
    - \delta_2 \emA \wedge \delta_1 \emE
   \end{eqtableau}
\end{equation}
As always in general relativity, the Hamiltonian takes the form of constraints
plus boundary terms.  The constraints consist of two Gauss law equations, one
for the self dual two form $\Sigma$ and the other for the electric field
$\emE$, together with the standard vector and scalar constraints.  When the
constraint equations are satisfied, the term at infinity equals $t^aP_a$ where
$P_a$ is the ADM 4-momentum.  In our signature, $t^aP_a$ is negative, so
$-t^aP_a = E^\ADM$, the standard ADM energy.  The equations of motion are just
Hamilton's equations:
\begin{equation}
    \delta\tilde{H} =\tilde\Omega(\delta,\, X_{\tilde{H}})\, .
\end{equation}
These are the field equations (\ref{eom2}) and (\ref{maxeq}) in a 3+1
form, expressed in terms of the canonical variables.

To conclude, we note two modifications which occur if the cosmological
constant is non-zero.  First, there is an extra term proportional to
$\Lambda \,\, {\rm Tr} [\Sigma \wedge \Sigma]$ in the action
(\ref{afAct}), where $\Lambda$ is the cosmological constant.  This
contributes a term proportional to $\Lambda {\rm Tr}
\sigma\wedge\sigma\wedge\sigma$ in the bulk term of the Hamiltonian
(\ref{AFPhSp}).  Second, the boundary conditions are modified.  If
$\Lambda$ is positive, it is natural to assume the Cauchy surfaces are
compact, whence there are no falloff conditions or boundary terms in
any of the expressions.  If $\Lambda$ is negative, the dynamical
fields approach asymptotically their values in the anti-de Sitter
space-time \cite{am}.

\section{Boundary Conditions}
\label{s3}

In this section, we specify the boundary conditions which define an
\textit{undistorted, non-rotating isolated horizon}.  As explained in the
Introduction, the purpose of these boundary conditions is to capture the
essential features of a non-rotating, isolated black hole in terms of the
intrinsic structure available at the horizon, without any reference to
infinity or to a static Killing field in space-time.  However, the boundary
conditions model a larger variety of situations.  For example, the
cosmological horizons in de Sitter space-times \cite{gh} are isolated
horizons, even though there is no sense in which they describe a black hole. 
As a result, the usual mechanics of cosmological horizons will be reproduced
within the framework of isolated horizons.  Other examples involve space-times
admitting gravitational and electro-magnetic radiation such as those described
in Section \ref{s3.2}.

The physical situation we wish to model is illustrated by the example
of figure \ref{exam}(a).  The late stages of the collapse pictured
here should describe a non-dynamical, isolated black hole.  However, a
realistic collapse will generate gravitational radiation which must
either be scattered back into the black hole or radiated to infinity. 
Physically, one expects most of the back-scattered radiation will be
absorbed rather quickly and, in the absence of outside perturbations,
the black hole will `settle down' to a steady-state configuration. 
This picture is supported by numerical simulations.  The continued
presence of radiation elsewhere in space-time however implies
$\mathcal{M}$ cannot be stationary.  As a result, the usual
formulations of black hole mechanics in terms of stationary solutions
cannot be easily applied to this type of physical black hole. 
Nevertheless, since the portion $\Delta$ of the horizon describes an
isolated black hole one would hope to be able to formulate the laws of
black hole mechanics in this context.  We will see in Sections 5 and 8
that this is indeed the case.

\subsection{Definition}
\label{s3.1}

We are now in a position to state our boundary conditions.  Since this
paper is concerned primarily with the mechanics of isolated horizons
in Einstein--Maxwell space-times, conditions on gravitational and
Maxwell fields are specified first and more general matter fields are
treated afterwards.

A \textit{non-rotating isolated horizon} is a sub-manifold $\Delta$ of
space-time at which the following five conditions hold:%
\footnote{Throughout this paper, $\hateq$ will denote equality at
points of $\Delta$.  For fields defined throughout space-time, a
single left arrow below an index will indicate the pull-back of that
index to $\Delta$, and a double arrow will indicate the pull-back of
that index to the preferred 2-sphere cross-sections $S_\Delta$ of
$\Delta$ introduced in condition \ref{bcKinem}.  For brevity of
presentation, slightly stronger conditions
were used in \cite{abf1}.}%
\begin{enumerate}
  \renewcommand{\theenumi}{\Roman{enumi}}
  \renewcommand{\labelenumi}{(\theenumi) }
  \renewcommand{\theenumii}{\alph{enumii}}
  \renewcommand{\labelenumii}{\theenumi\theenumii.\ }

\item\label{bcKinem} \textsl{$\Delta$ is topologically $S^2 \times \Re$
and comes equipped with a preferred foliation by 2-spheres
$S_\Delta$ and a ruling by lines transverse to those 2-spheres.}

These preferred structures give rise to a 1-form direction field $[n_a]$ and a
(future-directed) vector direction field $[\ell^a]$ on $\Delta$.  Furthermore,
any $n_a \in [n_a]$ is normal to a foliation of $\Delta$ by 2-spheres and we
further impose $dn \hateq 0$.  As a result, the equivalence class $[n_a]$ is
defined with respect to rescaling only by functions which are constant on each
$S_\Delta$.  A function with this property will be said to be
\textit{spherically symmetric}.  We `tie together' the normalizations of the
two direction fields by fixing $\ell^a n_a \hateq -1$ with $\ell^a \in
[\ell^a]$.  This leaves a single equivalence class $[\ell^a, n_a]$ of
direction fields subject to the relation
\begin{equation}\label{equivRel}
  (\ell^a, n_a) \sim (F^{-1} \ell^a, F n_a),
\end{equation}
where $F$ is any positive, spherically symmetric function on
$\Delta$.

\item\label{bcNull} \textsl{The soldering form $\tensor <\sigma_a^AA'>$ gives
rise to a metric in which $\Delta$ is a null surface with $[\ell^a]$ as its
null normal.}

So far the 1-forms $n_{a}$ are defined intrinsically on the 3-manifold
$\Delta$.  They can be extended uniquely to 4-dimensional, space-time 1-forms
(still defined only at points of $\Delta$) by requiring that they be null.  We
will do so and, for notational simplicity, denote the extension also by
$n_{a}$.  Given any null pair $(\ell^a, n_a)$ satisfying $\ell^a n_a \hateq
-1$, it is easy to show there exists a spin basis $(\iota^A, \o^A)$ satisfying
$|\iota^A \o_A| \hateq 1$ such that
\begin{equation}\label{nullBase}
\ell^a \hateq i \tensor <\sigma^a_AA'> \o^A \bar \o^{A'}
\qquad\mbox{and}\qquad n_a \hateq i \tensor <\sigma_a^AA'> \iota_A
\bar\iota_{A'}.
\end{equation}
We will work by \textit{fixing} a spin dyad $(\iota^A, \o^A)$ with
$\iota^A \o_A \hateq +1$ once and for all and regarding (\ref{nullBase})
as a condition on the field $\tensor <\sigma_a^AA'>$.  Finally, using
this spin dyad, we can complete $(\ell^a, n^a)$ to a null tetrad with
the vectors
\begin{equation}\label{fullBase}
  m^a \hateq i\tensor<\sigma^a_AA'> o^A \bar\iota^{A'}
  \qquad\mbox{and}\qquad \bar m^a \hateq i\tensor<\sigma^a_AA'>
  \iota^A \bar o^{A'}
\end{equation}
tangential to the 2-spheres $S_\Delta$.%
\footnote{Since the topology of $\Delta$ is $S^2 \times \Re$, the
tetrad vectors $m^a$ and $\bar{m}^a$ --- and the spin-dyad $(\i^A,
\o^A)$ --- fail to be globally defined.  Thus, when we refer to a
fixed spin-dyad, we mean dyads which are fixed on two patches and
related by a gauge transformation on the overlap.  Our loose
terminology is analogous to the one habitually used for (spherical)
coordinates on a 2-sphere.}

\item\label{bcGeom} \textsl{The derivatives of the spin dyad are
constrained by
\begin{equation}\label{derivCon}
\o^A \grad_{\pback{a}} \o_A \hateq 0 \qquad\mbox{and}\qquad
\iota^A \grad_{\pback{a}} \iota_A \hateq \mu \bar m_a,
\end{equation}
where $\mu$ is a real, nowhere vanishing, spherically symmetric function,
and $\grad_a$ denotes the unique torsion-free connection compatible
with $\tensor <\sigma_a^AA'>$}.  The function $\mu$ is one of the
standard Newman-Penrose spin coefficients.

\item\label{bcDynam} \textsl{All equations of motion hold at
$\Delta$}.  In particular:

\begin{enumerate}

\item\label{bcCompat} The $\rm SL(2,\Com)$ connection is compatible
with the soldering form: $D_a \lambda^A \hateq \grad_a \lambda^A$.

\item\label{bcEins} The Einstein equations hold: $G_{ab}+\Lambda g_{ab} \hateq 8\pi G
T_{ab}$, where $T_{ab}$ is the stress-energy tensor of the matter
fields under consideration and $G_{ab}$ is the Einstein tensor of the
metric compatible with $\tensor <\sigma_a^AA'>$.

\item\label{bcMaxEq} The electro-magnetic field strength $\emF$
satisfies the Maxwell equations: $d\emF \hateq 0$ and $d\dual\emF \hateq
0$.

\end{enumerate}

\item\label{bcMaxSym} \textsl{The Maxwell field strength $\emF$ has the
property that
\begin{equation}\label{EMCon}
\phi_1 :\hateq \half m^a \bar m^b (\emF - i\dual\emF)_{ab}
\end{equation}
is spherically symmetric}.
\end{enumerate}

These five conditions define a non-rotating isolated horizon.  Each is
imposed \textit{only} at the points of $\Delta$.  Let us now discuss
the geometrical and physical motivations behind the boundary
conditions, and see how they capture the intuitive picture outlined
above.

Conditions \ref{bcKinem} and \ref{bcNull} are straightforward.  The first is
primarily topological and fixes the kinematical structure of the horizon. 
(While the $S^{2}\times \Re$ topology is the most interesting one from
physical considerations, most of our results go through if $S^{2}$ is replaced
by a compact, 2-manifold of higher genus.  This issue is briefly discussed at
the end of Section \ref{s5}.)  The meaning of the preferred foliation of
$\Delta$ will be made clear in the discussion of condition \ref{bcGeom} below. 
The meaning of the preferred ruling can be seen immediately in condition
\ref{bcNull} which simply requires that $\Delta$ be a null surface, and
$[\ell^a]$ its null generator.  Thus, the preferred ruling singles out the
null generators of the horizon.

Condition \ref{bcDynam} is a fairly generic dynamical condition,
completely analogous to the one usually imposed at null infinity:
\textit{Any} set of boundary conditions must be consistent with the
equations of the motion at the horizon.  It is likely that this
condition can be weakened, e.g., by requiring only that the pull-backs
to $\Delta$ of the equations of motion should hold.  However, care
would be needed to specify the precise form of equations which are to
be pulled-back since pull-backs of two equivalent sets of equations
can be inequivalent.  We chose simply to avoid this complication.

Conditions \ref{bcKinem}, \ref{bcNull} and \ref{bcDynam} are weak; in
particular, they are satisfied by a variety of null surfaces in any
solution to the field equations.  It is condition \ref{bcGeom} which
endows $\Delta$ with the structure of an isolated horizon. 
Technically, this condition restricts the pull-back of the self-dual
connection compatible with $\tensor <\sigma_a^AA'>$.  (Note the
pull-backs to $\Delta$ are important because we have introduced the
dyad $(\i^A, \o^A)$ only at the surface itself.)  Geometrically, it is
equivalent to requiring the pairs $(\ell^a, n_a) \in [\ell^a, n_a]$ to
have the following properties:
\begin{enumerate}

\item $\ell^a$ is geodesic, twist-free, shear-free and
divergence-free.

\item $n^a$ is twist-free, shear-free, has nowhere vanishing, spherically
symmetric expansion
\begin{equation}\label{expnf}
\theta_{(n)} \hateq 2\mu.
\end{equation}
and vanishing Newman-Penrose spin coefficient $\pi :\hateq l^a\bar{m}^b
\nabla_a n_b$ on $\Delta$.

\end{enumerate}
As we will see in the next section, the only independent consequence of
condition \ref{bcGeom} for $\ell^a$ is its vanishing divergence; the rest then
follows from simple geometry.  The vanishing divergence of $\ell^a$ is
equivalent to the vanishing expansion of the horizon, We will see in Section
\ref{s4.1} that this implies there is no flux of matter falling across
$\Delta$, which in turn captures the idea that the horizon is isolated. 
Collectively, the consequences of condition \ref{bcGeom} for $n^a$ imply the
horizon is non-rotating and its intrinsic geometry is undistorted.  Requiring
$\mu$ to be nowhere vanishing is equivalent to requiring the expansion of
$n^a$ to be nowhere vanishing.  In black-hole space-times, we expect this
expansion to be strictly negative on future horizons and strictly positive on
past horizons.  On cosmological horizons, both signs are permissible.  In
section \ref{s6} we will impose further restrictions tailored to different
physical situations.  Finally, it is easy to verify that these conditions on
$n^a$ suffice to single out the preferred foliation of $\Delta$. 
\textit{Thus, we could have just required the existence of a foliation
satisfying the first three conditions, and used \ref{bcGeom} to conclude the
foliation is unique.}

Next, let us discuss the condition \ref{bcMaxSym} on the Maxwell
field.  At first sight, this requirement seems to be a severe
restriction.  However, if the Newman-Penrose component $\phi_0$ of the
electro-magnetic field, representing `the radiation field traversing
$\Delta$', vanishes in a neighborhood of $\Delta$ and boundary
conditions I through IV hold, then $\phi_1$ automatically satisfies
\ref{bcMaxSym}.  Heuristically, this feature can be understood as
follows.  {}From the definition of $\phi_1$ in (\ref{EMCon}), one can
see condition \ref{bcMaxSym} requires the electric and magnetic fluxes
across the horizon to be spherically symmetric.  If this condition did
not hold, one would intuitively expect a non-rotating black hole to
radiate away the asphericities in its electro-magnetic hair, giving
rise to a non-vanishing $\phi_0$.  Therefore, it is reasonable to
expect \ref{bcMaxSym} would hold once the horizon reaches equilibrium.

Let us now consider the generalization of the boundary conditions to
other forms of matter.  Conditions \ref{bcMaxEq} and \ref{bcMaxSym}
refer to the Maxwell field; the rest involve only the gravitational
degrees of freedom and are independent of the matter fields present at
the horizon.  We will now indicate how these two conditions must be
modified in the presence of more general forms of matter.  (For a
discussion of dilatonic couplings, see \cite{ack,30}.)  First, note
that condition \ref{bcDynam} is unambiguous.  Hence, \ref{bcMaxEq}
would simply be replaced with the field equations of the relevant
matter.  Condition \ref{bcMaxSym}, on the other hand, is more subtle. 
In fact, there is no universal analog of (\ref{EMCon}) which applies
to arbitrary matter fields; the boundary conditions used in place of
condition \ref{bcMaxSym} may vary from case to case.  There are,
however, two general properties which any candidate matter field and
its associated boundary conditions must possess:

\begin{enumerate}\setcounter{enumi}{4}
  \renewcommand{\theenumi}{\Roman{enumi}$^\prime$}
  \renewcommand{\labelenumi}{(\theenumi) }
  \renewcommand{\theenumii}{\alph{enumii}}
  \renewcommand{\labelenumii}{\theenumi\theenumii.\ }

\item\label{bcGenMat}
For any $(\ell^a, n_a) \in [\ell^a, n_a]$, a matter field must satisfy

\begin{enumerate}

\item\label{bcEnergy}
\textsl{The stress-energy tensor is such that
\begin{equation}\label{StrEnCon}
k^a \hateq -\tensor <T^a_b> \ell^b
\end{equation}
is \textit{causal}, i.e., future-directed, time-like or null}.

\item\label{bcFldSym} \textsl{The quantity
\begin{equation}\label{enDensCon}
    e :\hateq T_{ab} \ell^a n^b
\end{equation}
is spherically symmetric on $\Delta$}.

\end{enumerate}

\end{enumerate}

The first of these properties, \ref{bcEnergy}, is an immediate consequence of
the (much stronger) dominant energy condition which demands $- \tensor <T^a_b>
<k^b>$ be causal for \textit{any} causal vector $k^b$.  Like any energy
condition, this is a restriction on the \textit{types} of matter which may be
present near the horizon.  On the other hand, the property \ref{bcFldSym} is a
restriction on the possible \textit{boundary conditions} which may be imposed
on matter fields at the horizon.

We conclude this section with a remark on generalizations of these boundary
conditions.  Although our framework is geared to the undistorted,
non-rotating case, only the requirements on $n^a$ in condition \ref{bcGeom}
and the symmetry condition \ref{bcMaxSym} on the Maxwell field would have to
be weakened to accommodate distortion and rotation \cite{jl}.  Specifically,
it appears that, in presence of distortion, $\mu$ will not be spherically
symmetric and in presence of rotation $\pi$ will not vanish.  However, it
appears that a more general procedure discovered by Lewandowski will enable
one to introduce a preferred foliation of $\Delta$ and naturally extend the
present framework to allow for distortion and rotation.

Finally, in light of the sphericity conditions on $\theta_{(n)}$ and $\phi_1$,
one may be tempted to call our isolated horizons `spherical'.  However, the
Newman-Penrose curvature components $\Psi_3$ and $\Psi_4$ and the Maxwell
field component $\phi_2$ need not be spherical on $\Delta$ for our boundary
conditions to be satisfied.  Therefore, the adjectives `undistorted and
non-rotating' appear to be better suited to characterize our isolated
horizons.

\subsection{Examples}
\label{s3.2}

It is easy to check that all of these boundary conditions hold at the
event horizons of Reissner--Nordstr\"om black hole solutions (with or
without a cosmological constant).  Similarly, they hold at
cosmological horizons in de Sitter space-time.  Furthermore, if one
considers a spherical collapse, as in figure 2, they hold both on
$\Delta_1$ and $\Delta_2$ (at suitably late times).

In the non-rotating context, these cases already include situations
normally considered in connection with black hole thermodynamics. 
However, the isolated horizons in these examples are also Killing
horizons for globally defined, static Killing fields.  We will now
indicate how one can construct more general isolated horizons.

\begin{figure}
  \begin{center}\small
    \psfrag{Delta}{$\Delta$}
    \psfrag{n}{$n$}
    \psfrag{ell}{$\ell$}
    \psfrag{Psi00}{$\Psi_0 = 0$}
    \psfrag{Psi4}{$\Psi_{4}$}
    \psfrag{S}{$S$}
    \psfrag{calN}{$\mathcal{N}$}
    \includegraphics[height=4cm]{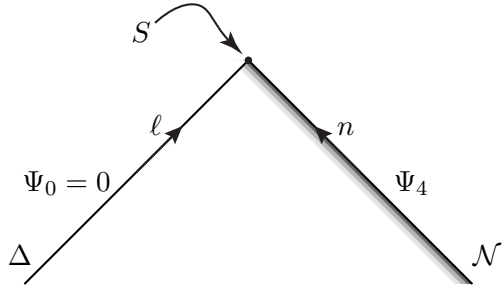}
    \caption{Space-times with isolated horizons can be
constructed by solving
    the characteristic initial value problem on two
intersecting null
    surfaces, $\Delta$ and ${\cal N}$.  The final solution
admits $\Delta$ as
    an isolated horizon.  Generically, there is radiation
arbitrarily close to
    $\Delta$ and no Killing fields in any neighborhood of
    $\Delta$.}\label{CIVP}
  \end{center}
\end{figure}

\begin{figure}
  \begin{center}\small
    \psfrag{fsI+}{$\scri^+$}
    \psfrag{fsH+}{$\mfs{H}^+$}
    \psfrag{calM}{$\mathcal{M}$}
    \psfrag{i+}{$i^+$}
    \psfrag{i0}{$\spi$}
    \psfrag{Delta}{$\Delta$}
    \psfrag{M}{$M$}
    \psfrag{Mp}{$M'$}
    \includegraphics[height=6cm]{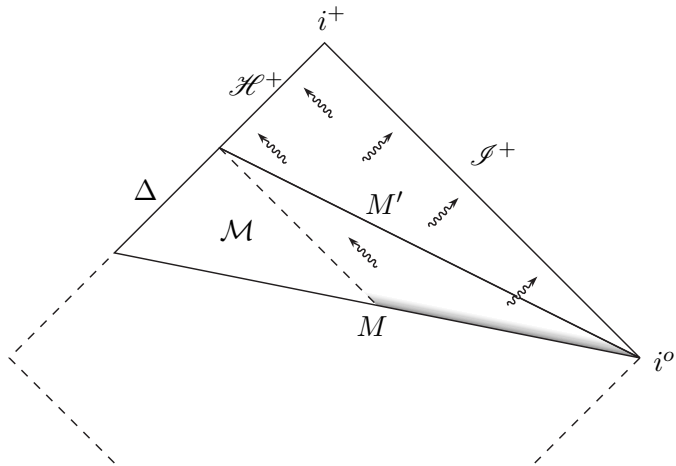}
    \caption{A space-time $\man$ with an isolated horizon
$\Delta$ as internal
    boundary and radiation field in the exterior can be
obtained by starting
    with an asymptotic region of Kruskal space-time and
modifying the initial
    data on the partial Cauchy surface $M$.  While the new
metric continues to
    be isometric with the Schwarzschild metric in a
neighborhood of $\Delta$,
    it admits radiation in a neighborhood of
infinity.}\label{exam1}
  \end{center}
\end{figure}

First, an infinite-dimensional space of examples can be constructed using
Friedrich's results \cite{hf}, and Rendall's extension \cite{r} thereof, on
the null initial value formulation (see figure 3).  In this framework, one
considers two null hypersurfaces $\Delta$ and $\mathcal{N}$ with normals
$\ell^a$ and $n^a$ respectively, which intersect in a 2-sphere $S$.  (At the
end of the construction $\Delta$ will turn out to be a non-rotating isolated
horizon.)  In a suitable choice of gauge \cite{hf}, the free data for vacuum
Einstein's equations consists of $\Psi_0$ on $\Delta$; $\Psi_4$ on
$\mathcal{N}$; and, the Newman Penrose coefficients $\lambda, \sigma, \pi,
\RePt\mu, \RePt\rho $ as well as the intrinsic metric ${}^2\!g_{ab}$ on the
2-sphere $S$.  Given these fields, there is a unique solution (modulo
diffeomorphisms) to the vacuum Einstein's equations in a neighborhood of $S$
bounded by (and including) the appropriate portions of $\Delta$ and
$\mathcal{N}$.  Let us set $\Psi_0 =0$ on $\Delta$ and $\lambda=\sigma =\rho=
\pi =0$ on $S$, $\mu = {\rm const}$ and ${}^2\!g_{ab}$ to be a (round)
2-sphere metric with area $\area$ on $S$.  Lewandowski \cite{jl} has shown
that, in the resulting solution, $\Delta$ is an isolated horizon with area
$\area$.  Note that in the resulting solution $\Psi_4$ need not vanish in any
neighborhood of $\Delta$, or, indeed, even \textit{on} $\Delta$.  Hence, in
the vacuum case, there is an infinite-dimensional space of (local) solutions
containing isolated horizons.  It turns out that, in this setting, there is
always a vector field $\xi^a$ in a neighborhood of $\Delta$ with $\xi^a \hateq
\ell^a$ and $\Lie_{\xi} g_{ab} \hateq 0$.  However, in general $\Lie_{\xi}
C_{abcd} \not\hateq 0$, where $C_{abcd}$ is the Weyl tensor of $g_{ab}$. 
Hence, in general, $\xi^a$ cannot be a Killing field of $g_{ab}$ even in a
neighborhood of $\Delta$.  (For details, see \cite{jl}.)  The
Robinson--Trautman solutions provide interesting examples of exact solutions
which bring out this point \cite{pc}: a sub-class of these solutions admit an
isolated horizon but no Killing fields whatsoever.  There is radiation in every
neighborhood of the isolated horizon but, in a natural chart, the metric
coefficients and several of their radial derivatives evaluated \textit{at}
$\Delta$ are the same as those of the Schwarzschild metric at its event
horizon.

A second class of examples can be constructed by starting with Killing
horizons and `adding radiation'.  To be specific, consider one
asymptotic region of the Kruskal space-time (figure 4) and a Cauchy
surface $M$ therein.  The idea is to change the initial data on this
slice in the region $r\ge 3m$, say, where $m$ is the Schwarzschild
mass of the initial space-time.  In the Einstein--Maxwell case, one can
use the strategy introduced by Cutler and Wald \cite{cw} in their
proof of existence of solutions with smooth null infinity.  In the
vacuum case, one can use the more general `gluing methods' recently
introduced by Corvino and Schoen \cite{schoen} to show that there
exists an infinite-dimensional space of asymptotically flat initial
data on $M$ which agree with the data for a Schwarzschild space-time
for $r<3m$ but in which the evolved space-time admits radiation. 
Using these methods, one would be able to construct `triangular
regions' ${\man}$ bounded by $M$, $\Delta$ and a partial Cauchy slice
$M'$ in the future of $M$.  If one takes ${\man}$ as the space-time of
interest, then $\Delta$ would serve as the isolated horizon at the
inner boundary.  Due to the presence of radiation, $\man$ will not
admit any global Killing field.  However, in a neighborhood of
$\Delta$, the 4-metric will be isometric to that of Schwarzschild
space-time.  Thus, in this case, there will in fact be four Killing
fields in a neighborhood of the isolated horizon.

The two constructions discussed above are complementary.  The first
yields more general isolated horizons but the final result is local. 
The second would provide space-times which extend from the isolated
horizon $\Delta$ to infinity but in which there is no radiation in a
neighborhood of $\Delta$.  We expect there will exist an
infinite-dimensional space of solutions to the vacuum Einstein
equations and Einstein--Maxwell equations which are free from both
limitations, i.e., which extend to spatial infinity and admit isolated
horizons with radiation arbitrarily close to them.  However, a
comprehensive treatment of this issue will be technically difficult. 
Given the current status of global existence and uniqueness results in
the asymptotically flat contexts, the present limitations are not
surprising.  Indeed, the situation at null infinity is somewhat
analogous: while the known techniques have provided several
interesting partial results, they do not yet allow us to show that
there exists a large class of solutions to Einstein's vacuum equations
which admit complete and smooth past and future null infinities,
$\scri^\pm$ and the standard structure at spatial infinity, $\spi$.

\section{Consequences of the Boundary Conditions}
\label{s4}

In this section, we will discuss the rich structure given to the
horizon by the boundary conditions.  The discussion is divided into
four subsections.  The first describes the basic geometry of an
isolated horizon.  The next two subsections examine the restrictions
on the space-time curvature and Maxwell field which arise from the
boundary conditions.  The last subsection contains a brief summary.

\subsection{Horizon Geometry}
\label{s4.1}

Let us begin by examining the consequences of condition III in the main
definition.  The condition on $\o_A$ immediately implies
\begin{equation}
\grad_{\pback{a}} \ell_b \hateq - 2U_{a} \ell_b
\end{equation}
for some 1-form $U_a$ on $\Delta$.  Hence, $\ell^a$ is geodesic, twist-free,
shear-free and divergence-free.  We will denote the acceleration of $\ell^a$
by $\tilde{\kappa}$:
\begin{equation}\label{kapDef}
\ell^a \grad_a \ell^b \hateq \tilde{\kappa} \ell^b,
\end{equation}
so that $\tilde{\kappa} \hateq -2\ell^a U_a$.  Note that $\tilde\kappa$ varies
with the rescaling of $\ell^a$.

Actually, the geodesic and the twist-free properties of $\ell^a$ follow
already from condition II which requires $\Delta$ to be a null surface with
$\ell^a$ as its null normal.  Furthermore, since $T_{ab}\ell^a\ell^b \ge 0$ by
condition \ref{bcEnergy}, and Einstein's equations hold at $\Delta$ (condition
IVb), we can use the Raychaudhuri equation
\begin{equation}\label{RayEq}
  \Lie_\ell \theta_{(\ell)}
   = -\half \theta_{(\ell)}^2 - \sigma_{ab} \sigma^{ab} +
   \omega_{ab} \omega^{ab} - R_{ab} \ell^a \ell^b
\end{equation}
to conclude the shear $\sigma_{ab}$ must vanish if the expansion
$\theta_{(\ell)}$ vanishes.  Thus, the only independent assumption contained
in the first of equations (\ref{derivCon}) in condition III is that $\ell^a$
is expansion-free, which captures the idea that $\Delta$ is an
\textit{isolated} horizon.  Finally, note that the Raychaudhuri equation also
implies that
\begin{equation}\label{RayRes}
R_{ab}\ell^a\ell^b \hateq 8\pi G T_{ab} \ell^a\ell^b \hateq 0
\end{equation}
i.e., that there is no flux of matter energy-momentum across the
horizon.

Let us now consider the properties of the vector field $n^a$.
The second equation in (\ref{derivCon}) implies
\begin{equation}
\grad_{\pback {a}}n_b \hateq (2U_an_b + 2\mu m_{(a} \bar m_{b)}),
\end{equation}
with $U_a \propto n_a$.  Hence, $n^a$ is twist-free, shear-free, has
spherically symmetric expansion, $2\mu$, and vanishing Newman-Penrose
coefficient $\pi \hateq \bar{m}^a \ell^b \grad_a n_b$.  The twist-free
property follows from the very definition of $n_a$, but the other three
properties originate in condition III of the Definition.

Next we turn to the intrinsic geometry of the horizon. Using the
definitions (\ref{fullBase}) of the co-vectors $m_a$ and $\bar m_a$,
one can show there exists a 1-form $\nu_a$ on $\Delta$ such that
\begin{equation}\label{nuDef}
  \pback{dm} \hateq -i \nu \wedge m   \qquad\mbox{and}\qquad
    \pback{d \bar m} \hateq i \nu \wedge \bar m.
\end{equation}
One consequence of these relations is that the Lie derivative along $\ell^a$
of the intrinsic metric on $\Delta$ must vanish: $\Lie_\ell g_{\pback{ab}}
\hateq 0$.  Thus, as mentioned in Section 3, the intrinsic geometry of an
isolated horizon is `time-independent'.  In particular, the Lie derivative
along $\ell^a$ of the volume form on the foliation 2-spheres $S_\Delta$
vanishes: $\Lie_\ell \tensor ^2<\epsilon> \hateq 0$.  Therefore, the areas of
all the $S_\Delta$ take the same value which we denote $\area$.  Finally, one
can show \cite{pr1} the scalar curvature ${}^2R$ of the 2-sphere cross
sections $S_\Delta$ is related to the 4-dimensional, Newman-Penrose curvature
scalars via: ${}^2R = -2\RePt{\Psi_2} + 2\Phi_{11} + R/12$.  In Section
\ref{s5}, it is shown that the quantity on the right side of this equation is
constant on $S_\Delta$.  Hence, the intrinsic metric $2m_{(a} \bar{m}_{b)}$ on
$S_\Delta$ is spherically symmetric.  In this sense, the horizon geometry is
undistorted.  However, the discussion of Section \ref{s3.2} shows spherical
symmetry will not extend, in general, to a neighborhood of $\Delta$.

\subsection{Form of the Curvature}
\label{s4.2}

Let us begin by exploring the effects of boundary condition \ref{bcGenMat} on
the form of the Ricci tensor.  Using the Raychaudhuri equation, we have
derived (\ref{RayRes}).  Whence $k^a$ in (\ref{StrEnCon}) must be proportional
to $\ell^a$.  Using the quantity $e$ defined in (\ref{enDensCon}), we then
have
\begin{equation}\label{strEnNull}
(8\pi G)^{-1} \left( R_{ab} \ell^b - \half R \ell_a + \Lambda \ell_a \right) 
\hateq T_{ab} \ell^b \hateq -e \ell_a,
\end{equation}
where $R$ is the scalar curvature.  This formula yields a series of
results for the Ricci tensor.  In terms of Newman-Penrose components
(\ref{NPPhi}), these read
\begin{equation}\label{zeroFlux}
  \Phi_{00} \hateq \Phi_{01} \hateq \Phi_{10} \hateq 0
        \qquad\mbox{and}\qquad
  \Phi_{11} + \frac{R}{8} - \frac{\Lambda}{2} \hateq4\pi G e.
\end{equation}
The first three results say, by way of the Einstein equation, there is
no flux of matter radiation falling through the isolated horizon.  The
fourth result implies the combination $\Phi_{11} + \frac{R}{8}$ is
spherically symmetric.

We can now explore the consequences of the condition III for the full
Riemann curvature.  Since any $SL(2,\Com)$-bundle over a 3-manifold is
trivializable, and since our 4-manifold $\man$ has the topology $M
\times \Re$ for some 3-manifold $M$, the connection $A$ can be
represented globally as a Lie algebra-valued 1-form.  Because of
(\ref{derivCon}), in the $(\iota, \o)$-basis, the pull-back to $\Delta$
of that connection must have the form
\begin{equation}\label{connForm}
\tensor <A_\pback{a}^AB> \hateq -(\tilde{\kappa} n_a + i \nu_a)
\iota^{(A} \o^{B)} - \mu \bar m_a \o^A \o^B.
\end{equation}
on $\Delta$.  (However, since the spin-dyad is defined only locally on
$\Delta$, the decomposition (\ref{connForm}) is also local.)  The
function $\mu$ appearing here is the same nowhere vanishing,
spherically symmetric function introduced in (\ref{derivCon}) and
$\tilde{\kappa}$ and $\nu$ are defined by (\ref{kapDef}) and
(\ref{nuDef}), respectively.  Now, using this expression for the
pull-back of the self-dual connection, we can simply calculate the
pull-back of its curvature to be
\begin{equation}\label{curvForm}
\pback{F}_{AB} \hateq-[d\tilde{\kappa} \wedge n + i d\nu]
\iota_{(A} o_{B)} +
[(\tilde{\kappa} \mu + \Lie_\ell \mu) n \wedge \bar m] o_A o_B,
\end{equation}
where we have suppressed all space-time indices for simplicity.  On
the other hand, we can also calculate the self-dual part of the
Riemann spinor in terms of Newman-Penrose components (see Appendix
\ref{curvApp}).  Then, using the compatibility of the self-dual
connection with the soldering form at the horizon, we get a second
expression for the pulled-back curvature:
\begin{equation}\label{curvPB}
  \begin{eqtableau}{1}
  \pback{F}_{AB} \hateq {} &\left[ \vphantom{\frac{R}{24}}
  \Phi_{00} n \wedge m + \Psi_0 n \wedge \bar m -
(\Psi_1 - \Phi_{01}) m \wedge \bar m \right] \iota_A \iota_B \\ - {}
&\left[ \Phi_{10} n \wedge m + \Psi_1 n \wedge \bar m -
\left( \Psi_2 - \Phi_{11} - \frac{R}{24} \right)
m \wedge \bar m \right] 2 \iota_{(A} o_{B)} \\ + {}
&\left[ \Phi_{20} n \wedge m + \left( \Psi_2 +
\frac{R}{12} \right)
 n \wedge \bar m - (\Psi_3 - \Phi_{21}) m \wedge
\bar m \right] o_A o_B.\\
  \end{eqtableau}
\end{equation}
Equating this expression with (\ref{curvForm}) one arrives at a series
of conclusions:

\begin{enumerate}

\item Since there is no $\iota_A \iota_B$ term in (\ref{curvForm}),
we find
\begin{equation}\label{Psizero}
  \Psi_0 \hateq \Psi_1 \hateq 0,
\end{equation}
where we have used (\ref{zeroFlux}).

\item Equating the $\i_{(A} \o_{B)}$ terms in the two expressions and
using (\ref{Psizero}) and (\ref{zeroFlux}) then yields
\begin{equation}\label{AbCurv}
n \wedge d\tilde{\kappa} \hateq 0 \qquad\mbox{and}\qquad
d\nu \hateq 2i \left( \Psi_2 - \Phi_{11} - \frac{R}{24} \right) m 
\wedge \bar m.
\end{equation}
The first expression here says the function $\tilde{\kappa}$ is
spherically symmetric.  In the second expression, the left side is
real and the quantity $i m \wedge \bar m \hateq \tensor ^2<\epsilon>$
on the right is real as well.  As a result, the coefficient $\Psi_2 -
\Phi_{11} - \frac{R}{24}$ must be real.  However, $\frac{R}{24}$ is
manifestly real and, due to its definition (\ref{NPPhi}), $\Phi_{11}$
is also real.  Thus, the second expression in (\ref{AbCurv}) implies
the imaginary part of $\Psi_2$ vanishes. This encodes the property 
that $\Delta$ is non-rotating.

\item Equating the remaining $\o_A \o_B$ terms similarly yields
$\Phi_{20} \hateq 0$ and $\Psi_3 \hateq \Phi_{21}$ as well as
\begin{equation}\label{kapCurv}
\Psi_2 + \frac{R}{12} \hateq \Lie_\ell \mu + \tilde{\kappa} \mu.
\end{equation}
Since $\mu$ and $\tilde\kappa$ are spherically symmetric, $\Psi_2 +
\frac{R}{12}$ must also have this property.
\end{enumerate}

In the later sections of this paper, we will consider the action and
phase space formulations of systems containing isolated horizons.  In
this discussion, it will be most useful to have a formula giving the
relations which arise from the boundary conditions among the
fundamental gravitational degrees of freedom in a simple, compact
form.  Using (\ref{curvPB}), the tetrad decomposition (\ref{SigComp})
of $\Sigma^{AB}$ and the above restrictions on the Newman-Penrose
curvature components, it is straightforward to demonstrate:
\begin{equation}\label{ActBC}
\pback{F}^{AB} \hateq \left [
\left( \Psi_2-\Phi_{11}- \frac{R}{24} \right) \delta^A_C \delta^B_D
-\left( \frac{3\Psi_2}{2}-\Phi_{11} \right) \o^A\o^B\i_C\i_D \right]
\pback{\Sigma}^{CD}.
\end{equation}
In the phase space formalism, only the pull-back to $S_\Delta$ of this
formula will be relevant.  This pull-back takes the simpler form
\begin{equation}\label{PhSpBC}
\pback[2]{F}^{AB} \hateq \left( {\Psi_2 - \Phi_{11} - \frac{R}{24}} 
\right) \, \pback[2]{\Sigma}^{AB}.
\end{equation}
Note that the essential content of this equation can be seen already
in (\ref{AbCurv}).

\subsection{Form of the Maxwell Field}
\label{s4.3}

The stress-energy tensor of a Maxwell field is given by
\begin{equation}\label{MaxStrEn}
\tensor <\emT_ab> `= \frac{1}{4\pi} \left[' <\emF_ac> <\emF_b^c>
 `- \tsfrac{1}{4}' <g_ab> <\emF_cd> <\emF^cd> \right].
\end{equation}
This stress-energy tensor satisfies the dominant energy condition and
hence, in particular, condition \ref{bcEnergy}.  Furthermore, one can
see from its definition that the trace of $\emT_{ab}$ is zero.

To see the restrictions which the boundary conditions place on the
Maxwell field, it is useful to recast this discussion in terms of
spinors as we did in the previous subsection for the self-dual
curvature.  The Maxwell spinor $\phi_{AB}$ is defined in terms of the
field strength via the relation
\begin{equation}
\emF_{ab} = \tensor <\sigma_a^AA'> <\sigma_b^BB'>
(\phi_{AB} \epsilon_{A'B'} + \epsilon_{AB} \bar\phi_{A'B'}).
\end{equation}
It is then straightforward to show that the stress-energy
(\ref{MaxStrEn}) can be expressed in terms of the Maxwell spinor as
\begin{equation}\label{SpinStrEn}
\tensor <\emT_ab> `= -\frac{1}{2\pi}' <\sigma_a^AA'> <\sigma_b^BB'>
<\phi_AB> <\bar\phi_A'B'>.
\end{equation}
We have already seen in the previous subsections that the general
matter field conditions, \ref{bcGenMat}, and the Raychaudhuri equation
imply the stress energy tensor must satisfy (\ref{strEnNull}).  Using
the spinorial definition (\ref{nullBase}) of $\ell^a$, this
restriction on the stress-energy tensor gives two important results
for the Maxwell field:
\begin{equation}\label{MaxSpinRes}
\phi_0 :\hateq \phi_{AB} \o^A \o^B \hateq 0 \qquad\mbox{and}\qquad
\eme \hateq \frac{1}{2\pi} \phi_{AB} \i^A \o^A \cdot \bar \phi_{A'B'}
\bar\i^{A'} \bar\o^{B'} \hateq \frac{1}{2\pi} \abs{\phi_1}^2.
\end{equation}
Here, $\eme$ is the quantity $e$ introduced in (\ref{enDensCon})
specialized to the Maxwell field.  The second equation shows the
spherical symmetry of $\phi_1$ we imposed in (\ref{EMCon}) does indeed
guarantee the spherical symmetry of $\eme$ required by
(\ref{enDensCon}).  However, the remaining Newman-Penrose component of
the Maxwell field, $\phi_2$, is completely unconstrained.  In
particular, it need not be spherically symmetric.

Since $\phi_1$ is spherically symmetric, we can express it in terms of
the electric and magnetic charges contained within the horizon.  To do
so, consider the general form of the Maxwell field compatible with
(\ref{MaxSpinRes}):
\begin{equation}\label{MaxTensRes}
\emF \hateq \ell \wedge (\phi_2 m + \bar\phi_2 \bar m) - 2 \left(
\ImPt{\phi_1}
\tensor^2<\epsilon> - \RePt{\phi_1} \dual\tensor^2<\epsilon> \right).
\end{equation}
Here, $\tensor ^2<\epsilon>$ denotes the volume form on $S_\Delta$ and
$\dual \tensor ^2<\epsilon>$ denotes its (space-time) dual.  Now,
since $\tensor ^2<\epsilon>$ is defined with respect to the
\textit{inward} pointing unit space-like normal (see Appendix
\ref{Conv}), we have
\begin{eqset}\label{Charge}
  -4\pi Q_\Delta &\hateq \oint_{S_\Delta} \dual\emF \hateq
    -2 \oint_{S_\Delta} \RePt{\phi_1}\tensor ^2<\epsilon> \\
\interject{and}
  -4\pi P_\Delta &\hateq \oint_{S_\Delta} \emF \hateq
    -2 \oint_{S_\Delta} \ImPt{\phi_1}\tensor ^2<\epsilon>, \\
\end{eqset}
where $Q_\Delta$ and $P_\Delta$ denote the electric and magnetic
charges contained \textit{within} $S_\Delta$.  Since $\phi_1$ is
spherically symmetric, its real and imaginary parts can be pulled
outside the integrals and we calculate it to be
\begin{equation}\label{NPCharge}
  \phi_1 \hateq \frac{2\pi}{\area} (Q_\Delta + iP_\Delta).
\end{equation}
Here, $Q_\Delta$ and $P_\Delta$ are naturally spherically symmetric,
but may as yet vary from one $S_\Delta$ to another.  However, the
remaining boundary condition on the Maxwell field, (\ref{bcMaxEq}),
requires the Maxwell equations hold at the horizon.  The Maxwell
equations pulled-back to $\Delta$ applied to the field strength
(\ref{MaxTensRes}) with $\phi_1$ given by (\ref{NPCharge}) imply the
charges $Q_\Delta$ and $P_\Delta$ must be constant over the entire
horizon $\Delta$.  It should be noted that this constancy is caused by
\textit{boundary conditions} and not by equations of motion in the
bulk.  As a result, $Q_\Delta$ and $P_\Delta$ are constant over
$\Delta$ in any history satisfying our boundary conditions and not
just `on-shell'.

\subsection{Summary}
\label{s4.4}

As we have seen in this section, the boundary conditions place many
restrictions on both the gravitational and electro-magnetic degrees of
freedom.  We will collect the results we have found here.  These
results use not only the boundary conditions, but also the fact that
the only form of matter we consider is a Maxwell field.

The Newman-Penrose components of the Maxwell field at the horizon are
constrained by
\begin{equation}
  \phi_0 \hateq 0  \qquad\mbox{and}\qquad
  \phi_1 \hateq \frac{2\pi}{\area} (Q_\Delta+iP_\Delta).
\end{equation}
The remaining component, $\phi_2$, is an arbitrary complex function
over $\Delta$.

The Newman-Penrose components of the Ricci tensor are
\begin{equation}
  R \hateq 4\Lambda \qquad\mbox{and}\qquad \Phi_{ij} \hateq 2G
  \phi_i \bar\phi_j \quad\mbox{with}\quad i,j = 0,1,2.
\end{equation}
The second equation is simply a consequence of the Einstein equation
and (\ref{SpinStrEn}).  The Newman-Penrose components of the Weyl
tensor satisfy
\begin{equation}
  \Psi_0 \hateq \Psi_1 \hateq 0  \qquad\mbox{and}\qquad
  \Psi_3 \hateq \Phi_{21}.
\end{equation}
The component $\Psi_4$ is an arbitrary complex function over $\Delta$.
The remaining component, $\Psi_2$, is related to the acceleration
$\tilde{\kappa}$ of $\ell^a$ and the expansion $2\mu$ of $n^a$ via:
\begin{equation}\label{psi2prime}
\Psi_2 + \frac{\Lambda}{3} \hateq \Lie_\ell \mu + \tilde{\kappa}
\mu,
\end{equation}
where $\Lambda$ is the cosmological constant.  It follows in
particular that $\Psi_2$ is real and spherically symmetric.  Finally,
these components also satisfy
\begin{equation} \label{dnu}
d\nu = 2 \left( \Psi_2 - \Phi_{11} - \frac{\Lambda}{6} \right)
\tensor ^2<\epsilon>,
\end{equation}
where $\nu$ is a connection on the frame bundle of $S_\Delta$ and
$\tensor ^2<\epsilon>$ is its volume form.

\section{Surface Gravity and the Zeroth Law}
\label{s5}

The zeroth law of black hole mechanics states that the surface gravity
$\kappa$ of a stationary black hole is constant over its horizon.  In
subsection \ref{RedGauge}, we will extend the standard definition of
$\kappa$ to arbitrary non-rotating isolated horizons using only
structure available at the horizon.  A key test of the usefulness of
this definition comes from the zeroth law: Does the structure of
$\Delta$ enable us to conclude $\kappa$ is constant on $\Delta$
without reference to a static Killing field?  In Section \ref{s5.2} we
will show the answer is in the affirmative.  Thus, our more
general definition of surface gravity is consistent with our notion of
isolation of the horizon.  Furthermore, we will see that the structure
of $\Delta$ is rich enough to enable us to express $\kappa$ in terms
of the parameters $r_\Delta$, $Q_\Delta$ and $P_\Delta$ of the
isolated horizon.

\subsection{Gauge reduction and surface gravity}
\label{RedGauge}

Our set-up suggests we define surface gravity using the acceleration,
$\tilde{\kappa}$, of the vector field $\ell^a$.  However, since the
acceleration fails to be invariant under the rescalings of $\ell^a$, we need
to normalize $\ell^a$ appropriately.  As mentioned in the Introduction, the
usual treatments of black hole mechanics in static space-times accomplish this
by identifying $\ell^a$ with the restriction to the horizon of that static
Killing field which is unit at infinity.  For a generic isolated horizon,
there will be no such Killing field and our procedure can only involve
structures defined \textit{on} the horizon.  Since $\ell^a$ is null, and its
expansion, twist and shear vanish, we cannot hope to normalize it by fixing
one of its own geometric characteristics.  However, the normalizations of
$\ell^a$ and $n^a$ are intertwined and we \textit{can} hope to normalize $n^a$
by fixing its expansion.  The normalization of $\ell^a$ will then be fixed.

To implement this strategy, let us begin by examining the available
gauge freedom.  The correspondence (\ref{nullBase}) between the fixed
spin dyad $(\i^A, \o^A)$ and the preferred direction fields $[\ell^a,
n_a]$ breaks the original $SL(2,\Com)$ internal gauge group at
$\Delta$ to $\Com U(1)$.  However, as we shall see shortly, the
residual gauge invariance has a somewhat unusual structure.  To see
the nature of the residual gauge, it is simplest temporarily to
consider a fixed soldering form and a variable spin dyad.

The most general transformation of the spin dyad which preserves the
correspondence (\ref{nullBase}) is
\begin{equation}\label{actGauge}
  (\iota^A, o^A) \mapsto \left( e^{\Theta - i\theta} \iota^A,
    e^{-\Theta + i\theta} o^A \right).
\end{equation}
Here, $\Theta$ and $\theta$ are both real functions over $\Delta$.
Under these residual gauge transformations, one can show the null
tetrad transforms as
\begin{equation}\label{tetGauge}
  \begin{eqtableau}{2}
    \ell^a &\mapsto e^{-2\Theta} \ell^a &\qquad
      n^a &\mapsto e^{2\Theta} n^a \\
    m^a &\mapsto e^{2i\theta} m^a &\qquad
      \bar m^a &\mapsto e^{-2i\theta} \bar m^a.\\
   \end{eqtableau}
\end{equation}
Thus, $\Theta$ accounts for the rescaling freedom in $\ell^a$ and $n_a$ which
must be eliminated to define the surface gravity unambiguously.  $\Theta$ is
restricted to be spherically symmetric by (\ref{equivRel}).  On the other
hand, $\theta$ is arbitrary and allows a general transformation on the frame
bundle of each $S_\Delta$.  Furthermore, the function $\mu$ appearing in
(\ref{derivCon}) is not gauge invariant, but transforms according to
\begin{equation}\label{expnGauge}
  \mu \mapsto e^{2\Theta} \mu.
\end{equation}
Note, however, that the spherical symmetry and nowhere-vanishing
property of $\mu$ are preserved by this transformation.  Finally, the
fields $\tilde{\kappa}$ and $\nu$ introduced in Section \ref{s4.1}
transform as
\begin{equation}\label{connGauge}
\tilde{\kappa} \mapsto e^{-2\Theta} (\tilde{\kappa} - 2 \Lie_{\ell} \Theta)
\qquad\mbox{and}\qquad  \nu_a \mapsto \nu_a - 2 \grad_a \theta.
\end{equation}
The transformation of $\tilde{\kappa}$ is the usual one for the
acceleration of a vector field when that field is rescaled, and
preserves the spherical symmetry of $\tilde{\kappa}$.  Meanwhile, the
transformation of $\nu_a$ is that of a $\rm U(1)$ connection.

We are now ready to fix the normalization of $\ell^a$.  The strategy
outlined in the beginning of this subsection can be implemented
successfully thanks to the following three non-trivial facts.  First,
the expansion of the properly normalized $n^a$ in a
Reissner--Nordstr\"om solution is `universal': irrespective of the
mass, electric and magnetic charges or cosmological constant, on the
future, outer black hole event horizons in these solutions, $\theta_{(n)}
\hateq -2/r_0$ where $r_0$ is the radius of the horizon. 
Motivated by this and the relation $\theta_{(n)} = 2\mu$ (see
(\ref{expnf})) we are led to set%
\footnote{To accommodate cosmological horizons, we will have to allow $\mu$ to
be strictly positive.  This issue will be discussed in section \ref{s6}.  For
now, we focus on black hole horizons and assume $\mu$ is strictly negative. 
Modifications required to accommodate positive $\mu$ are straightforward.}
\begin{equation}\label{fixReal}
  \mu \hateq -\frac{1}{\rad}
\end{equation}
on a \textit{generic}, non-rotating, Einstein--Maxwell isolated horizon
of radius $\rad$ (i.e., of area $\area = 4\pi r_\Delta^{2}$).  Second, the
gauge-freedom (\ref{tetGauge}) available to us is such that we can
\textit{always} achieve the normalization (\ref{fixReal}) of $\mu$. 
Third, it is obvious from (\ref{expnGauge}) that this condition
exhausts the freedom in $\Theta$ completely.  In particular,
therefore, in a single stroke, this procedure fixes the normalization
of $n^a$ (and $\ell^a$) \textit{and} gets rid of the awkwardness in
the gauge freedom.  The restricted gauge freedom is now given simply
by:
\begin{equation}\label{tetGauge2}
  \begin{eqtableau}{2}
    \ell^a &\mapsto  \ell^a &\qquad
      n^a &\mapsto  n^a \\
    m^a &\mapsto e^{2i\theta} m^a &\qquad
      \bar m^a &\mapsto e^{-2i\theta} \bar m^a,\\
   \end{eqtableau}
\end{equation}
where $\theta$ is an arbitrary real function on $\Delta$; the local
gauge group is reduced to $U(1)$.  We can now return to our usual
convention wherein the spin dyad is fixed while the soldering form
varies.  The true residual gauge transformations are then the duals of
(\ref{actGauge}), with $\Theta = 0$, acting on the soldering form. 
The effect of these residual transformations on the null tetrad and
the other fields discussed here remain the same as those in
(\ref{tetGauge2}).

From now on, we will assume that $n^a$ and $\ell^a$ are normalized via
(\ref{fixReal}), denote the resulting acceleration of $\ell^a$ by
$\kappa$ and refer to it as the \textit{surface gravity of the
isolated horizon} $\Delta$.  By construction, our general definition
reduces to the standard one in Reissner--Nordstr\"om solutions.

To conclude this subsection, let us consider the gauge freedom in
Maxwell theory.  We just saw that a partial gauge fixing of the
$SL(2,\Com)$ freedom in the gravitational sector is necessary to
define the surface gravity in the absence of a static Killing field. 
The situation with the electric and magnetic potentials is analogous. 
In conventional treatments \cite{heus} these can be defined using the
global static Killing field which, however, is unavailable for a
generic isolated horizon.  The idea again is to resolve this problem
through a partial gauge fixing.  Since the only available parameters
are the radius $r_\Delta$ and the charges $Q_\Delta, P_\Delta$,
dimensional considerations suggest the electric potential $\Phi \hateq
\emA_a\ell^a$ is proportional to $Q_\Delta/r_\Delta$ on the horizon. 
We fix the proportionality factor using the standard value of $\Phi$
in Reissner--Nordstr\"om solutions.  Thus, in the general case, we
partially fix the gauge by requiring
\begin{equation}\label{fixMax}
\ell^a \emA_a \hateq \Phi :\hateq \frac{Q_\Delta}{\rad}.
\end{equation}
It turns out that this strategy of gauge-fixing also makes the
variational principle well-defined for the Maxwell action.

The situation with the magnetic potential, however, is more subtle. 
Since there is no obvious expression for the magnetic potential in
terms of $\emA_a$, we cannot formulate a definition similar to
(\ref{fixMax}).  Instead, we will appeal to the well-known duality
symmetry of the Maxwell field.  Thus, in the remainder of the paper,
we set $P_\Delta = 0$ \textit{in the main discussion}.  The results
for isolated horizons with both electric and magnetic charge will
follow from those including only electric charge by a duality
rotation.

\subsection{Zeroth law}
\label{s5.2}

We already know from (\ref{AbCurv}) that $\kappa$ is spherically
symmetric.  Therefore, to establish the zeroth law, it only remains to
show that $\Lie_\ell \kappa$ also vanishes.  Recall first from
(\ref{ActBC}) that on $\Delta$, $\pback{F}^{AB}$ is given by:
\begin{equation}
\pback{F}^{AB} \hateq \left [
\left( \Psi_2-\Phi_{11}- \frac{R}{24} \right) \delta^A_C \delta^B_D
-\left( \frac{3\Psi_2}{2}-\Phi_{11} \right) \o^A\o^B\i_C\i_D \right]
\pback{\Sigma}^{CD}.
\end{equation}
Let us now consider the Bianchi identity $\pback{D F}^{AB} =0$.  Transvecting
it with $\i_A\o_B$, we obtain
\begin{equation}\label{bi}
\Lie_\ell\, \left( \Psi_2 - \Phi_{11} - \frac{R}{24} \right) \hateq 0.
\end{equation}
In the Einstein--Maxwell system, $\Phi_{11} \hateq 2G|\phi_1|^2$ and
$\phi_1$ is constant on $\Delta$ (see (\ref{NPCharge})).  Similarly,
since the stress-energy tensor is trace-free, $R = 4\Lambda$ is also a
constant.  Hence it follows that $\Lie_\ell \Psi_2 \hateq0$.  Finally,
(\ref{psi2prime}) and our gauge condition (\ref{fixReal}) immediately
imply:
\begin{equation}\label{psi2}
  \kappa= -r_\Delta \left( \Psi_2 +\frac{\Lambda}{3} \right).
\end{equation}
Hence, we conclude $\Lie_\ell \kappa \hateq 0$, as desired.  This
establishes the zeroth law of the mechanics of isolated horizons in
the Einstein--Maxwell theory.

We will now obtain an explicit expression for $\kappa$ in terms of the
parameters of this isolated horizon.  The key fact is that the first
Chern number of the pull-back to $S_\Delta$ of the connection $\nu$
introduced in (\ref{nuDef}) is two.  This can be seen in a number of
ways, but it is essentially equivalent to the Gauss-Bonet theorem for
a 2-sphere because $\pback[2]{\nu}$ can be identified with a $SO(2)$
connection in the frame bundle of $S_\Delta$.  Using this property, we
have
\begin{equation}\label{Chern}
2 = \frac{-1}{2\pi i} \oint_{S_\Delta} id\nu =
\frac{-1}{2\pi} \oint_{S_\Delta}
2 \left[ \left( \Psi_2 + \frac{R}{12} \right) - \left( \Phi_{11} +
\frac{R}{8} \right) \right] \cdot
\tensor ^2<\epsilon>,
\end{equation}
where we have used (\ref{AbCurv}) in the last equality here.  Now, we
have seen in (\ref{kapCurv}) that the first term in the brackets here
is spherically symmetric, and in (\ref{zeroFlux}) that the second term
is proportional to the quantity $e$ which has been restricted to be
spherically symmetric.  As a result, the entire integrand on the right
side of \ref{Chern} can be pulled through the integral.  The remaining
integral simply gives the area $\area$ of $S_\Delta$.  Thus,
\begin{equation}
\Psi_2 - \Phi_{11} -\frac{R}{24} \hateq - \frac{2\pi}{\area}
\end{equation}
Let us now specialize to the Einstein--Maxwell case.  Then, $R =
4{\Lambda}$ and $\Phi_{11}= (G/2 r_\Delta^4) (Q_\Delta^2+P_\Delta^2)$. 
Therefore, using (\ref{psi2}) we can express surface gravity in terms
of $r_\Delta$, $Q_\Delta$ and $P_\Delta$:
\begin{equation}\label{surfGrav}
\kappa = \frac{1}{2\rad} \left( 1 - \frac{G
(Q_\Delta^2+P_\Delta^2)}{r_\Delta^2}
-\Lambda r_\Delta^2 \right).
\end{equation}

We conclude with a few remarks.

\newcommand{\rem}[1]{\medskip\noindent\hbox to
0pt{#1.\hss}\indent\ignorespaces}

\rem{1} The final expression (\ref{surfGrav}) for $\kappa$ is formally
identical to the expression for the surface gravity of a
Reissner--Nordstr\"om black hole in terms of its radius and charge. 
This may be surprising since we have not restricted ourselves to
static situations.  However, if it is possible to express $\kappa$ in
terms of the parameters $r_\Delta$, $Q_\Delta$, $P_\Delta$ and
$\Lambda$ alone, this agreement \textit{must} hold if the general
expression is to reduce to the standard one on event horizons of
static black holes.  In our treatment, the agreement can be
technically traced back to our general strategy for fixing the
normalization of $\ell^a$.

\rem{2} It may also be surprising that, although we do not have a
static Killing field at our disposal, it was possible to define
$\kappa$ unambiguously and it turned out to satisfy the zeroth law. 
Furthermore, we could express $\kappa$ in terms of the parameters of
the isolated horizon, irrespective of the details of gravitational and
electro-magnetic radiation outside the horizon.  This was possible
because of two facts.  First, the boundary conditions could
successfully extract just that structure from static black holes which
is relevant for these thermodynamical considerations.  Second, at its
core, the zeroth law is really local to the horizon; it does not know,
nor care, about the space-time geometry away from the horizon. 
Physically, this meets one's expectation that the degrees of freedom
of a black hole in equilibrium should `decouple' from the excitations
present elsewhere in space-time.

\rem{3} Our expression (\ref{kapCurv}) of surface gravity in terms of Weyl and
scalar curvature is universal, i.e., holds independent of the particular
matter sources so long as they satisfy the mild energy condition
(\ref{StrEnCon}).  Furthermore, in all these cases, $\kappa$ is spherically
symmetric and the Bianchi identity ensures (\ref{bi}).  The restriction to
Maxwell fields as the only source has been used here only to show that
$\frac{1}{4 \pi G} \Lie_{\ell} \left( \Phi_{11} +\frac{R}{8} -
\frac{\Lambda}{2} \right) = \Lie_{\ell} T_{{ab}}\ell^{a}n^{b} \equiv
\Lie_{\ell} e$ vanishes on $\Delta$.  Thus, in the present setting of
non-rotating isolated horizons, the zeroth law would hold for more general
matter provided its stress energy tensor satisfies this last restriction. 
This condition is satisfied, for example, by dilatonic matter \cite{ack,30}.

\rem{4} In the main definition, we assumed $\Delta$ has topology $S^{2}\times
\Re$.  If $S^{2}$ is replaced by a compact 2-manifold of higher genus, the
results of Section \ref{s4} and the proof of constancy of $\kappa$ on $\Delta$
remain unaffected.  However, in obtaining the explicit expression
(\ref{surfGrav}) of surface gravity, we used the Gauss-Bonnet theorem.  Hence
this expression is not universal but depends on the genus of $S_\Delta$.

\section{Action and Hamiltonian}
\label{s6}

As pointed out in the Introduction, to arrive at an appropriate
generalization of the the first law of black hole mechanics, we first
need to define the mass of an isolated horizon.  The idea is to arrive
at this definition through Hamiltonian considerations.  In Section
\ref{s6.1}, we introduce an action principle which yields the correct
equations of motion despite the presence of the internal boundary
$\Delta$.  In Section \ref{s6.2}, we pass to the Hamiltonian theory by
performing a Legendre transform and in \ref{s6.3} we discuss
Hamilton's equations of motion. We will find that, due
to the form of the boundary conditions, there are subtle differences between
the Lagrangian and the Hamiltonian frameworks because the latter allows more
general variations than the former.  

As in previous sections, in the main discussion we assume the space-time under
consideration is asymptotically flat with vanishing magnetic charge and
discuss at the end the modifications required to incorporate non-zero
$\Lambda$ and $P_\Delta$.

\subsection{Action}
\label{s6.1}

Recall from Section \ref{s2} that, in absence of internal boundaries,
the action for Einstein--Maxwell theory is given by:
\begin{equation} \label{action1}
\tilde{S}(\sigma,A,\emA) =\frac{-i}{8\pi G} \int_\man \Tr [\Sigma \wedge
F] +\frac{1}{8 \pi} \int_\man \emF \wedge \dual \emF +\frac{i}{8 \pi G}
\int_{C_\infty} \Tr [\Sigma \wedge A]
\end{equation}
In the presence of internal boundaries, however, this action need not
be functionally differentiable.  This is the case with our present
boundary conditions at $\Delta$.  In \cite{ack}, the required
modifications were discussed for the case when all histories $(\sigma,
A, \emA)$ under consideration induce a fixed area $\area$ and
electro-magnetic charges $Q_\Delta$ (and $P_\Delta$) on $\Delta$ and a
differentiable action was obtained by adding to $\tilde{S}$ a surface
term at $\Delta$.  (For a general discussion of surface terms in the
action, see \cite{js}.)

The strategy of fixing the parameters of the isolated horizon was appropriate
in \cite{ack} because the aim of that analysis was to provide the classical
framework for entropy calculations of horizons with {\it specific} values of
their parameters.  In this paper, on the other hand, we need a Hamiltonian
framework which is sufficiently general for the discussion of the first law,
in which one must allow the parameters to change.  Therefore, we now need to
allow histories with all possible values of parameters.  Note, however, that a
key consequence of the boundary conditions is that the area $\area$ and charge
$Q_\Delta$ of the isolated horizon are constant in time.  Therefore, the
values of these parameters are fixed in any one history (although they may
vary from one history to another).  Now, since all the fields are kept fixed
on the initial and final surfaces in the variational principle and the values
of our dynamical fields on either of these surfaces determine $\area$ and
$Q_{\Delta}$ for that history, one is only allowed to use those variations for
which $\delta\area$ and $\delta Q_{\Delta}$ vanish identically.  This fact
simplifies the task of finding an appropriate action considerably: For
example, as far as the action principle is concerned, one can continue to use
the action used in \cite{ack}.

There is however, a further subtlety which will lead us to use a more
convenient boundary term in the action.  Because of the nature of the
variational principle discussed above, we are free to add \textit{any}
function of the horizon parameters $\area$ and $Q_\Delta$ without affecting
the Lagrangian equations of motion.  In the framework considered in
\cite{ack}, this just corresponds to the freedom of adding a constant (with
appropriate physical dimension) to the Lagrangian.  However, since the full
class of histories now under consideration allows arbitrary areas $\area$ and
charges $Q_\Delta$, the freedom is now more significant: it corresponds to
changing the Lagrangian by a \textit{function} of parameters $\area,
Q_\Delta$.  While the variational procedure used to derive the Lagrangian
equations of motion is completely insensitive to this freedom, the
Hamiltonians resulting from these Lagrangians will clearly be different.  Can
all these Hamiltonians yield consistent equations of motion?  The answer is in
the negative.  The reason lies in subtle differences between the Lagrangian
and Hamiltonian variations.  In the Hamiltonian framework, the phase-space is
based on a fixed space-like 3-surface $M$.  Since the values of $\area$ and
$Q_\Delta$ can vary from one history to another, they can also vary from one
point of the phase space to another.  In obtaining Hamilton's equations,
$\delta H = \Omega(\delta, X_H)$, we must now allow phase space tangent
vectors $\delta$ which {\it can} change the values of parameters $\area,
Q_\Delta$.  Consequently, with boundary conditions such as ours, the burden on
the Hamiltonian is greater than that on the Lagrangian.  It turns out that
these additional demands on the Hamiltonian suffice to eliminate the apparent
functional freedom in its expression.  More precisely, the requirement that
Hamilton's equations of motion be consistent for \textit{all} variations
$\delta$ in the phase space, including the ones for which $\delta\area$ and
$\delta Q_\Delta$ do not vanish, determine the Hamiltonian completely.  (There
is no freedom to add a constant because, with only Newton's constant $G$ and
the speed of light $c$ at our disposal, there is no constant function on the
phase space with dimensions of energy.)  One can then work backwards and
single out the expression of the action, which, upon Legendre transform,
yields the correct Hamiltonian.  We will follow this strategy.

\begin{figure}
  \begin{center}\small
    \psfrag{Delta}{$\Delta$}
    \psfrag{i0}{$\spi$}
    \psfrag{M1}{$M_1$}
    \psfrag{M2}{$M_2$}
    \psfrag{S1}{$S_1$}
    \psfrag{S2}{$S_2$}
    \psfrag{calM}{$\mathcal{M}$}
    \includegraphics[height=4cm]{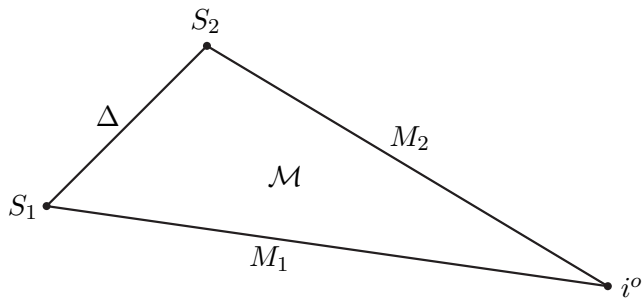}
\caption{Region $\man$ of space-time considered in the variational
principle is bounded by two partial Cauchy surfaces $M_1$ and $M_2$. 
They intersect the isolated horizon $\Delta$ in preferred 2-spheres
$S_1$ and $S_2$ and extend to spatial infinity $\spi$.}\label{HamPic}
\end{center}
\end{figure}

Fix a region of space-time whose inner boundary is an undistorted,
non-rotating isolated horizon, as depicted in figure 5.  This region $\man$ is
bounded in the past and future by space-like hypersurfaces $M_1$ and $M_2$
respectively which intersect the horizon $\Delta$ in preferred 2-sphere
cross-sections $S_1$ and $S_2$ and extend to spatial infinity $\spi$.  Since
the space-time $\man$ is asymptotically flat at spatial infinity, the fields
obey the standard falloff conditions at $\spi$.  The interior boundary
$\Delta$ is a non-rotating isolated horizon which satisfies the boundary
conditions discussed in section 3.  It turns out that, to obtain a
well-defined action principle, we need to impose an additional condition at
$\Delta$:
\begin{equation} \label{restriction}
\oint_{S_{\Delta}} (\nu\cdot \ell)\,\, {}^{2}\!\epsilon = 0 
\end{equation}
for any (preferred) 2-sphere cross-section $S_{\Delta}$ of $\Delta$,
where ${}^2\!\epsilon$ is the natural alternating tensor on $S_\Delta$
(see Appendix A.2).  Note that this restriction is \textit{very} mild
since it only asks that the spherically symmetric part (with respect
to ${}^{2}\!\epsilon$) --- or, equivalently, the zero mode --- of
$A\cdot \ell$ be real on $\Delta$.  Then, the required action is given
by:
\begin{equation} \label{action}
\begin{eqtableau}{1}
S &=\frac{-i}{8\pi G} \int_\man \Tr [\Sigma \wedge F]
+\frac{1}{8 \pi} \int_\man \emF \wedge \dual \emF
+\frac{i}{8\pi G} \int_{C_\infty} \Tr [\Sigma \wedge A]
+\frac{1}{8\pi G} \int_{\Delta} (r_\Delta\, \Psi_2)\, 
{}^\Delta\!\epsilon ,
\end{eqtableau}
\end{equation}
where ${}^\Delta\!\epsilon$ is the volume form on $\Delta$ compatible with our
normalization for $\ell^a$ (see Appendix A).  In Section \ref{s6.3}, we find
the resulting Hamiltonian does yield a consistent set of equations.  That
discussion will also show that the term added in the passage from
(\ref{action1}) to (\ref{action}) is uniquely determined by the consistency
requirement.

Note that (\ref{action}) does not have the Chern-Simons boundary term
at $\Delta$ used in \cite{ack}.  However, if one restricts oneself to
histories with a fixed value of $a_{\Delta}$ as in \cite{ack},
(\ref{action}) is completely equivalent to the action used there. 
(The mild restriction (\ref{restriction}) was not discussed in
\cite{ack} but is needed also in the action principle used there.)  In
particular, as we will see in Section \ref{s6.2}, the symplectic
structure obtained from the present action (\ref{action}) again has a
boundary term at $\Delta$.  If one works with a fixed $\area$, this
term reduces to the Chern-Simons symplectic structure of \cite{ack}. 
For non-perturbative quantization \cite{6,abk}, it is this
symplectic structure that plays the important role; the form of the
action is not directly relevant.  

In this paper we have chosen to work with (\ref{action}) because it is more
convenient for the Hamiltonian framework with variable $\area$, needed in the
discussion of the first law.  Furthermore, this form of the action appears to
extend naturally to isolated horizons with distortion and rotation and also
may be better suited for quantization in these more general contexts.

\subsection{Hamiltonian Framework}
\label{s6.2}

To pass to the Hamiltonian framework, we need to perform the Legendre
transform.  As in Section \ref{s2.2}, we begin by introducing the necessary
structure on the 4-manifold $\man$.  First, foliate $\man$ by partial Cauchy
surfaces $M$ with the following properties: i) the 3-manifolds $M$ intersect
the horizon at the preferred 2-spheres $S_\Delta$ such that the unit time-like
normal $\tau^a$ to them coincides with the vector $(\ell^a+n^a)/\sqrt{2}$ at
$S_\Delta$; and, ii) they extend to spatial infinity and intersect $C_\infty$
in 2-spheres $S_\infty$.  Next, fix a smooth time-like vector field $t^a$,
transverse to the leaves $M$ and a function $t$ such that: i) $t^a \nabla_a
t=1$ on $\man$; ii) $t^a$ tends to the unit time translation
\textit{orthogonal} to $M$ at spatial infinity; and, iii) $t^a$ tends to
$\ell^a$ on the horizon $\Delta$.  (The restriction on $t^{a}$ that it be
orthogonal to the foliation at infinity has been made only for simplicity and
can be removed easily by suitably modifying the discussion of the physical
interpretation of surface terms in the Hamiltonian.)  Finally, we will
restrict ourselves to physically interesting situations in which $M$ are
partial Cauchy surfaces for the space-time region $\man$ under consideration
and adapt our orientations to the case in which the projection of $n^{a}$ into
$M$ is a radial vector which points \textit{away} from the region $\man$. 
(See figure 6 and the discussion that follows in Section \ref{s6.3}).

The 3+1 decomposition of the space-time fields can now be performed exactly as
in equations (\ref{3+1def}) and (\ref{3+1field}).  Once again, the phase space
consists of quadruples ($\tensor<A_a^AB>$,
$\tensor<\Sigma_ab^AB>$,$\emA_a$,$\emE_{ab}$) on the 3-manifold $M$ satisfying
appropriate boundary conditions.  The conditions at infinity are the same as
before, namely (\ref{afbc}).  However, there are additional boundary
conditions at the horizon.  First, because of (\ref{connForm}), the form of
the gravitational connection $A$ is restricted at $S_\Delta$:
\begin{equation}
\pback[2]{A}^{AB} \hateq -i\pback[2]{\nu} \,\i^{(A}\o^{B)} +
\frac{1}{\rad} \,\bar{m}\, \o^A\o^B \, .
\end{equation}
Next, the curvature of $A$ is restricted by (\ref{PhSpBC}) and (\ref{Chern})
to satisfy $\pback[2]{F} \hateq -2\pi\pback[2]{\Sigma}/\area$, and the
electro-magnetic field $\emF$ is restricted by (\ref{NPCharge}).  Finally, the
requirement that the action principle be well-defined imposes the mild
restriction (\ref{restriction}) at $\Delta$.
 
The Legendre transform is straightforward using the procedure outlined in
Section \ref{s2.2}.  The only new element is the treatment of boundary terms
at the horizon which requires the use of the boundary conditions listed above. 
In order to state the final result, we have to introduce some notation.  The
space of our connections $\nu$ on $\Delta$ has the structure of an affine
space.  Let us choose any one of these connections $\puto\nu$, satisfying
$\puto\nu \cdot \ell =0$ as the `origin'.  (For example, $\puto\nu$ can be the
`static magnetic monopole potential' on every $S_{\Delta}$.)  Then using
boundary conditions, it is easy to show that any other connection $\nu$ can be
expressed as:
\begin{equation} \label{psi}
\nu = \puto\nu + \eta + d\psi 
\end{equation}
where the 1-form $\eta$ on $\Delta$ satisfies: $\ell\cdot \eta =0$, $\ell\cdot
d\eta =0$ and $d\, {}^{\star}\eta =0$ where the dual is taken under the metric
on each $S_{\Delta}$.  Note that there is the obvious gauge freedom $\psi
\mapsto \psi + {\rm const}$ in the choice of the function $\psi$.  Let us
eliminate it by requiring
\begin{equation} \label{restriction:psi}
\oint_{S_{\Delta}} \psi\, {}^2\!\epsilon= 0 \end{equation}
on any $S_{\Delta}$. Then the Legendre transform of the action $S$ 
of (\ref{action}) is given by:
\begin{equation} \label{legendre}
S(\sigma, A, \emA) =  \int dt\, \left[\frac{i}{8\pi G}\,\int_{M(t)}
\Lie_{t} A \wedge \Sigma - \frac{-i}{8\pi G}\, \oint_{S_{\Delta}(t)}
\Lie_{\ell} \psi \, {}^{2}\!\epsilon \right] -\int dt H_t,
\end{equation}
with
\begin{equation}
H_t = \tilde H_t - \oint_{S_\Delta} \left( \frac{\rad}{4\pi G}\, \Psi_2 -
\frac{Q_\Delta}{2\pi\rad}\phi_1 \right) \, {}^2\!\epsilon,
\end{equation}
where $\tilde H_t$ is defined in (\ref{AFPhSp}).  Thus, the
Hamiltonian has the familiar form: As in (\ref{AFPhSp}), the bulk term
is a volume integral of constraints weighted by Lagrange multipliers
determined by $t^a$ and the surface term at infinity gives
$t^aP^\ADM_a = -E^\ADM$.  However, now there is now a surface term at
the horizon as well.  

In order to bring out the similarity and differences in the two
surface terms, let us express the term at infinity using the Weyl
curvature.  Assuming the field equations hold near infinity and with
the shift $\vec N$ set to zero at $S_\infty$ because of our current
assumptions on the asymptotic value of $t^a$, we have \cite{am3}
\begin{equation}\label{Ham}
H_t = \int_M \mbox{constraints}
+\lim_{r_o\rightarrow \infty}\oint_{S_{r_o}} \left( \frac{r_o}
{4\pi G}\,\Psi_2 \right) \,{}^2\!\epsilon
- \oint_{S_\Delta} \left( \frac{\rad}{4\pi G}\, \Psi_2 -
\frac{Q_\Delta}{2\pi\rad}\phi_1 \right) \tensor^2<\epsilon>
\end{equation}
Thus, only the `Coulombic parts' of the two curvatures enter the
expressions of the two surface terms.  However, while the surface term at
infinity depends \textit{only} on the gravitational curvature, the term
at the horizon depends also on the Maxwell curvature.

The symplectic structure also acquires a term at the
horizon.
\begin{equation}\label{Sym}
\Omega (\delta_1, \delta_2) = \tilde\Omega (\delta_1, \delta_2)
-\frac{i}{8\pi G}\int_{S_{\Delta}} [\delta_{1}\psi\,\delta_{2}
\,({}^{2}\!\epsilon) - \delta_{2}\psi \,\,\delta_{1}\, 
({}^{2}\!\epsilon)]
\end{equation}
where $\tilde\Omega$ is defined in (\ref{AFPhSp}).  (As mentioned in Section
\ref{s6.1}, if we restrict ourselves to the phase space corresponding to a
fixed value of $a_{\Delta}$, the surface term in (\ref{Sym}) reduces to the
Chern-Simons symplectic structure for the connection $\pback[2]{A}$ on
$S_{\Delta}$).  The presence of the surface term in the symplectic structure
is rather unusual.  For instance, although there is a boundary term in the
action at infinity, the symplectic structure does not acquire a corresponding
boundary term.  Also note that we have not added new `surface degrees of
freedom' at the horizon (in contrast with, e.g., \cite {12,13}).  Indeed, our
phase space consists only of the standard bulk fields on $M$ which normally
arise in Einstein--Maxwell theory (see Section \ref{s2.2}) and whose values on
$S_\Delta$ are determined by their values in the bulk by continuity.  If
anything, the boundary conditions \textit{restrict} the degrees of freedom on
$\Delta$ by relating fields which are independent in the absence of
boundaries.  The symplectic structure on the space of these bulk fields simply
acquires an extra surface term.  In the classical theory, while this term is
essential for consistency of the framework, it does not play a special role. 
In the description of the quantum geometry of the horizon and the entropy
calculation \cite{6,abk}, on the other hand, this term turns out to be
crucial.

\subsection{Hamilton's Equations}
\label{s6.3}

Hamilton's equations are
\begin{equation}\label{HamEq}
  \delta H_t = \Omega(\delta,\, X_H),
\end{equation}
where $X_H$ is the Hamiltonian vector field associated with the given time
evolution vector field $t^a$ and $\delta$ is an arbitrary variation of the
fields.  Unlike in the discussion of the action, all fields appearing in the
Hamiltonian are defined only on the space-like surface $M$.  Hamilton's
equations describe the time evolution of these fields.  In particular, there
is no a priori reason to expect the area or charge of the isolated horizon to
be constant in time.  The constancy of the area and charge must arise, if at
all, as equations of motion of the theory.  As we already noted, since the
linearized fields $\delta$ in (\ref{HamEq}) can have $\delta\area \not=0$ and
$\delta Q_\Delta\not=0$, there are `more' Hamilton's equations than what one
would have naively thought from the Lagrangian perspective.  The question is
whether the additional equations ensure the area and charge are conserved and
if the full set of equations is self-consistent.

Let us summarize the consequences of Hamilton's equations for the
Hamiltonian and symplectic structure introduced in the last
subsection.  The bulk equations of motion give the standard
Einstein--Maxwell equations expressed in terms of connection variables. 
As usual, the variation of the term at infinity in the Hamiltonian
cancels the surface term at infinity arising from the variation of the
bulk term.  Also, the equations of motion preserve the boundary
conditions at infinity.

Thus, it only remains to examine the horizon terms.  Using the relation
(\ref{psi2}) among $\Psi_2$, $\kappa$ and $\rad$, and equating the horizon
terms on the two sides of Hamilton's equations, we obtain:
\begin{equation} \label{hor:eom}
{}^{2}\!\dot\epsilon = 0  \quad {\rm and} \quad \dot{\psi} = 
\nu\cdot\ell\, .
\end{equation}
The first of these equations in particular guarantees that the horizon
area does not change under time-evolution.  The second equation
follows from (\ref{psi}) which defines $\psi$ and is thus a
consistency condition.  Note also that the restriction
(\ref{restriction:psi}) on $\psi$ is preserved in time because of
(\ref{restriction}). 

Finally, it is natural to ask whether Hamilton's equations imply
$\dot{Q}_\Delta = 0$.  The answer is in the affirmative.  However,
this result is a consequence of a bulk equation of motion which
guarantees $\oint_S \dot{\emE}=0$ where $S$ is any closed two surface. 
If we take $S=S_{\Delta}$, the obvious consequence is
\begin{equation}
  \dot{Q}_\Delta = 0.
\end{equation}

To summarize, there exists a unique consistent Hamiltonian formulation
in the presence of inner boundaries which are isolated horizons.  The
symplectic structure is given by (\ref{Sym}), and the Hamiltonian by
(\ref{Ham}).  The bulk equations of motion are the standard 3+1
versions of the Einstein--Maxwell equations.  There are, however,
additional equations on the horizon 2-sphere $S_\Delta$ which
guarantee that $\rad$ is a constant of motion.

For simplicity, in the main discussion we restricted ourselves to zero
magnetic charge and cosmological constant, only one asymptotic region
and only one inner boundary.  However, these restrictions can be
easily removed.  If there is more than one asymptotic region and/or
isolated horizon inner boundary, one need only include surface terms
for each of these boundary 2-spheres.  The incorporation of a non-zero
magnetic charge and cosmological constant has a slightly more
significant effect.  As in Section \ref{s2.2}, the presence of a
cosmological constant changes the boundary conditions and the surface
terms at infinity.  The symplectic structure is unchanged.  But, as
discussed in Section \ref{s7.1}, the surface term at the horizon in
the expression of the Hamiltonian acquires additional terms involving
$P_\Delta$ and $\Lambda$.

\section{Physics of the Hamiltonian}
\label{s7}

In this section, we will examine the expression (\ref{Ham}) of the Hamiltonian
in some detail and extract physical information from it.  In Section
\ref{s7.1}, we will argue that the surface term at $S_{\Delta}$ should be
identified with the mass of the isolated horizon.  In Section \ref{s7.2}, we
will show that the numerical value of the Hamiltonian in any static solution
vanishes identically so that the mass of the isolated horizon in such a
space-time reduces to the ADM mass at infinity.  Finally, in Section
\ref{s7.3}, we will argue that, in any solution to the field equations which
is asymptotically flat at null and spatial infinity and asymptotically
Schwarzschild at future time-like infinity, the Hamiltonian (\ref{Ham}) equals
the total energy radiated away through future null infinity, $\scri^+$.  In
these space-times, the mass of the isolated horizon then equals the future
limit of the Bondi mass, exactly as one would intuitively expect.  We should
emphasize, however, that the argument rests on assumptions on the asymptotic
behavior of various fields (particularly near $i^+$) and we do \textit{not}
prove the existence of solutions to field equations with this behavior. 
Therefore, the discussion of Section \ref{s7.3} has a different status from
the rest of the paper.  Its primary purpose is to strengthen our intuition
about the Hamiltonian and the mass of the isolated horizon.

\subsection{Isolated Horizon Mass}
\label{s7.1}

For any physical system, energy can be identified with the numerical, on-shell
value of the generator of the appropriate time translation.  In Minkowskian
field theories, for example, it is the generator of motions on phase space
which correspond to space-time diffeomorphisms along a constant time-like
vector field.  Consider, as a second example, the theory of gravitational and
electro-magnetic radiation in general relativity.  In this case, one can
construct a phase space of radiative modes at null infinity and the total
radiated energy is the numerical value of the Hamiltonian generating a time
translation in the BMS group at null infinity \cite{jmp,as}.  Finally, in the
physics of fields which are asymptotically flat at spatial infinity, the ADM
energy arises as the on-shell, numerical value of the Hamiltonian generating
an asymptotic time-translation.  If the space-time under consideration admits
several asymptotic regions (as, for example, in the Kruskal picture) then the
energy in any one asymptotic region is given by the generator of a
diffeomorphism which is an asymptotic time-translation in the region under
consideration and asymptotically identity in all other regions.

These considerations suggest we define the energy associated with a given
isolated horizon $\Delta$ to be the numerical, on-shell value of the generator
of a diffeomorphism which is a time translation at $\Delta$ and asymptotically
identity.  To obtain an expression for this energy, let us examine the
expression (\ref{Ham}) of the Hamiltonian $H_t$.  The bulk term vanishes on
shell and the term at infinity does not contribute if the vector field $t^a$
vanishes at spatial infinity.  Thus, the required expression is given simply
by the surface term at $S_\Delta$.  Furthermore, since the vector field $t^a$
tends to $\ell^a$ on $\Delta$ and, by construction, $\ell^a$ defines the
`rest-frame' of the isolated horizon, this energy can be identified with the
mass $M_\Delta$ of $\Delta$.  Thus, the Hamiltonian considerations lead us to
set
\begin{equation}
M_\Delta = -\oint_{S_{\Delta}} \, \left[ \frac{r_\Delta}{4\pi G}\,
\left( \Psi_2 +\frac{\Lambda}{3} \right) -
\frac{Q_\Delta-iP_\Delta}{2\pi r_\Delta}\, \phi_1 \right] \tensor^2<\epsilon>,
\end{equation}
where we have now allowed for a non-zero cosmological constant
$\Lambda$ and magnetic charge $P_\Delta$.

For purposes of the first law, it will be useful to rewrite this expression
by eliminating the curvatures $\Psi_2$ and $\phi_1$ in favor of surface
gravity $\kappa$ and electro-magnetic scalar potential $\Phi$.  Using
(\ref{NPCharge}), (\ref{fixMax}) and (\ref{psi2}), when $P_\Delta= 0$,
we have:
\begin{equation} \label{mass}
 M_\Delta = \frac{1}{4\pi G}\, \kappa\, a_\Delta + \Phi\, Q_\Delta
\end{equation}
Note that the expression is formally identical with the familiar Smarr formula
\cite{smarr} for the mass of a Reissner--Nordstr\"om black hole.  One also
knows directly (i.e., without making an appeal to black hole uniqueness
theorems) that the ADM mass of {\it any} static black hole in the
Einstein--Maxwell theory is given by (\ref{mass}) \cite{heus}.  Thus, as with
our definition of surface gravity $\kappa$, although $M_\Delta$ is defined
using only the structure at the isolated horizon $\Delta$, it agrees with the
standard definition of black hole mass for static solutions.  The reason
behind this agreement will become clear in the next sub-section.  However, in
general (non-static) space-times, due to the presence of radiation, the ADM
mass at infinity is quite distinct from the isolated horizon mass $M_\Delta$. 
When constraints are satisfied, we have
\begin{equation}\label{HamVal}
  H_t = M_\Delta - E^\ADM \, ,
\end{equation}
and we will see in Section \ref{s7.3} that the numerical value of $- H_t$ can
be identified with the total energy radiated through future null infinity.
Finally, note that $M_{\Delta}$ has a specific contribution from the Maxwell
field.  We will see that this contribution is rather subtle but quite crucial
to adequately handle charged processes in the first law.  As far as we are
aware, none of the general, quasi-local expressions of mass contain this
precise contribution from the Maxwell field.  Thus, in the charged case, it
appears that $M_{\Delta}$ does not agree with any of the proposed quasi-local
mass expressions.

A natural question is whether $M_{\Delta}$ is positive.  Let us first consider
the case with zero cosmological constant.  Then, by fixing the value of the
charge $Q_{\Delta}$ and minimizing $M_{\Delta}$ with respect to $\rad$, one
finds $M_{\Delta}$ is not only positive, but also bounded below:
$M^{2}_{\Delta} \ge G Q_{\Delta}^{2}$.  At the minimum, i.e. when
$M_{\Delta}^{2} = GQ_{\Delta}^{2}$, the surface gravity $\kappa$ vanishes. 
However, unlike $M_{\Delta}$, $\kappa$ can be negative.  This structure is
familiar from Reissner--Nordstr\"om solutions, where the same inequality holds,
$\kappa$ vanishes at extremality, is positive on the outer horizon and
negative on the inner.  However, it was not obvious that this entire
structure would remain intact on general isolated horizons.

\begin{figure}
  \begin{center}\small
    \psfrag{i}{$r=\infty$}
    \psfrag{z}{$r=0$}
    \psfrag{M}{$M$}
    \psfrag{a}{a}
    \psfrag{b}{b}
    \psfrag{c}{c}
    \psfrag{d}{d}
    \includegraphics[height=6cm,width=12cm]{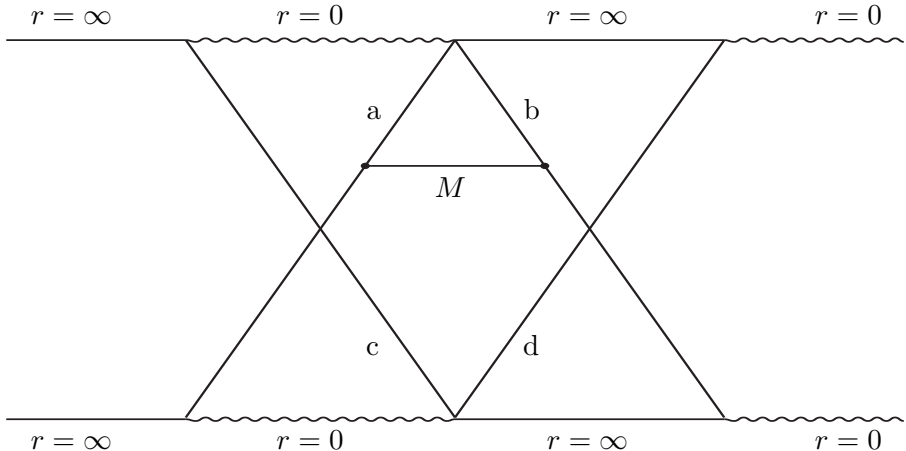} 
    \caption{Schwarzschild--de Sitter space-time.  Surface (a) is black hole
    horizon, (b) and (d) are cosmological horizons, and (c) is a white hole
    horizon.  The isolated horizon boundary conditions are satisfied on all
    four.  However, the Hamiltonian framework is tailored to future horizons
    of types (a) and (b) which are boundaries of space-time regions admitting
    partial Cauchy surfaces $M$.  The expansion of $n^a$ (and hence $\mu$) is
    negative on (a) and positive on (b).  Surface gravity $\kappa$ and mass
    $M_\Delta$ are positive on both (a) and (b).}\label{Cosmo}
  \end{center}
\end{figure}

Let us now consider the case when the cosmological constant $\Lambda$ is
non-zero.  If $\Lambda$ is negative, $M_{\Delta}$ is again positive and
$M^{2}_{\Delta} > GQ^{2}_{\Delta}$.  In this case, is natural to impose
asymptotically anti-de Sitter boundary conditions, whence one only expects
`black-hole type' horizons.  If $\Lambda$ is positive, one also has
cosmological horizons and the situation becomes more involved.  The resulting
complications are illustrated by the Schwarzschild-de Sitter space-time (see
figure 6).  The Hamiltonian framework is physically useful only in those
situations in which $M$ is a partial Cauchy surface for the space-time region
$\man$ under consideration.  The isolated horizons in this case are future
boundaries of space-time such as (a) and (b) in the figure.  Surface gravity
as well as mass are positive on the black hole horizon (a).  The case of the
cosmological horizon requires a reconsideration of the sign conventions we
previously adopted (see footnote 4).  Specifically, in the construction of the
Hamiltonian framework of Section \ref{s6.1}, we chose our orientations by
assuming the projection of $n^{a}$ into $M$ is `outward pointing' at
$S_\Delta$ relative to $M$.  With this choice, the expansion of $n^{a}$ (and
hence the Newman--Penrose coefficient $\mu$) is negative on the black hole
horizon (a), but positive on the cosmological horizon (b).  Since we assumed,
again for definiteness, that $\mu$ is negative in Sections \ref{s5} and
\ref{s6}, certain trivial modifications are needed to accommodate cosmological
future horizons of the type (b).  With these changes, the surface gravity and
mass are again positive on (b).

To summarize, it is future horizons of type (a) and (b) that are of physical
interest in our Hamiltonian framework.  For them, the surface gravity and mass
are positive in Schwarzschild--de Sitter space-time and we expect the
situation to be similar for general isolated horizons with positive
cosmological constant.  More detailed considerations suggest that the
interpretation of $\area$ as entropy and $\kappa$ as temperature are
applicable only to such horizons.

\subsection{Static solutions}
\label{s7.2}

The phase space now under consideration admits a 2-parameter family
of static solutions, labeled by $M$ and $Q$ --- the Reissner--Nordstr\"om
solutions.  Let us begin by evaluating the Hamiltonian $H_t$ (of (\ref{Ham}))
on these solutions using for $t^a$ the static Killing field.  Then, the volume
integrals will vanish since the constraints are satisfied and only
contributions from $S_{\infty}$ and $S_{\Delta}$ will remain.  The term at
infinity equals the negative of the ADM mass while, as noted above, the
horizon term is given by $\frac{\kappa\area}{4 \pi G} +\Phi Q_\Delta$.  Now,
it is well known from the Smarr formula that on a Reissner--Nordstr\"om
space-time, the value of the ADM mass is precisely $\frac{\kappa\area}{4 \pi
G} +\Phi Q_\Delta$.  Therefore the value of the Hamiltonian $H_t$ when
evaluated on static space-times is zero.  (The same reasoning extends to the
case when $\Lambda$ and $P_\Delta$ are non-zero.)

This feature is not accidental; there is a general argument from
symplectic geometry which `explains' this vanishing of $H_t$.  We
will conclude this sub-section by presenting the argument.

In symplectic geometry, Hamilton's equations are $\delta H = \Omega(\delta,
X_H)$, where $\delta$ is an arbitrary variation and $X_H$ is the Hamiltonian
vector field.  A stationary solution (such as a Reissner--Nordstr\"om solution
in the sector of Einstein--Maxwell theory now under consideration) is one at
which the Hamiltonian vector field either vanishes or generates pure gauge
evolution.  In either case, the symplectic structure evaluated on $X_H$ and
any other vector field $\delta$ vanishes.  Therefore, at these points of the
phase space, $\delta H = 0$ for \textit{any} variation $\delta$.  In
particular $\delta H=0$ for variations relating two nearby stationary
solutions.  Let us suppose the phase space is such that the space of
stationary solutions is connected (an assumption satisfied by the
Reissner--Nordstr\"om family in our case).  Then, the Hamiltonian must take
some constant `preferred value' on all stationary solutions.  Now, let us
suppose there is no natural energy scale in the theory.  (This assumption is
satisfied in our case because $M_\Delta$ and $Q_\Delta$ are not fixed on our
phase space and because one cannot construct a quantity with dimensions of
mass from $G$ and $c$ alone.)  Then, this `preferred value' must be zero.

\subsection{Hamiltonian equals Radiative Energy}
\label{s7.3}

We now present a result which provides an intuitive interpretation of the
Hamiltonian $H_t$ and a further motivation for our definition of the isolated
horizon mass $M_\Delta$.  More precisely, using suitable regularity
assumptions, we will show that, when the equations of motion (with $\Lambda
=0$) are satisfied and $t^a$ is adapted to the natural rest frames at the
horizon, the numerical value of $H_t$ equals $t^aP^\Rad_a = -\ERad$, where
$\ERad$ is flux of energy radiated across $\scri^+$.  However, as
explained in
the beginning of Section \ref{s7}, because we will need a number of new
assumptions, the considerations of this sub-section are not as self-contained
as those of the rest of the paper.

Let us assume that the underlying space-time $\man$ is of the type indicated
in figure (1.a).  That is, we assume i) the space-time is asymptotically flat
at future null infinity $\scri^+$ and asymptotically Schwarzschild at
time-like infinity $i^+$; ii) the Bondi news tensor on $\scri^+$ tends to zero
as one approaches $\spi$ and $i^+$; iii) the isolated horizon $\Delta$ extends
to $i^+$; and, iv) the boundary of $\man$ consists of $\Delta$, $\scri^+ \cup
i^+ \cup \spi$ and a partial Cauchy surface $M$.

Fix a conformal completion $(\hat{\man}, \hat{g}_{ab})$, of $(\man, g_{ab})$
which has $\scri^+$ as its (future) null boundary.  Appendix C summarizes the
structure available at $\scri^+$.  Let us begin by recalling that part of its
structure which we will need in this subsection.  The conformal factor
$\Omega$ vanishes at $\scri^+$ and $\hat{n}_a := \hat\nabla_a\Omega$ is the
null normal to $\scri^+$.  The conformally rescaled metric $\hat{g}_{ab}$
induces a degenerate metric $\hat{q}_{ab}$ at $\scri^+$ which satisfies
$\hat{q}_{ab} \hat{V}^a = 0$ on $\scri^+$ if and only if the tangent vector
$\hat{V}^a$ to $\scri^+$ is proportional to $\hat{n}^a$.  Although
$\hat{q}_{ab}$ is degenerate, we can define its `inverse' $\hat{q}^{ab}$ via
$\hat{q}_{ab}\hat{q}^{bc}\hat{q}_{cd}= \hat{q}_{ad}$.  This `inverse' is
unique up to additions of terms of the form $\hat{n}^{(a} \hat{V}^{b)}$ for
some vector field $\hat{V}^a$ tangent to $\scri^+$.  The phase space of
radiative modes of the gravitational and electro-magnetic fields at $\scri^+$
can be coordinatized%
\footnote{More precisely, the radiative phase space consists of certain
equivalence classes of connections on $\scri^+$.  Thus it has a natural affine
space structure.  In introducing this coordinatization, an equivalence class
of `trivial' connections has been chosen as the origin.  The fact that the
phase space is an affine --- rather than a vector --- space has some subtle
but important consequences.  These will be ignored here as they do not affect
the issues now under discussion.  For details, see \cite{jmp,as}.}
by pairs of fields $(\gamma_{ab},\emA_a)$ defined intrinsically on $\scri^+$.
The fields $\gamma_{ab}$ code the gravitational degrees of freedom; they are
symmetric, transverse (i.e., $\gamma_{ab}n^b=0$) and trace-free (i.e.,
$\gamma_{ab}\hat{q}^{ab}=0$).  These properties imply that $\gamma$ has
precisely two independent components which represent the two radiative modes
of the gravitational field.  The Maxwell degrees of freedom are coded in the
1-form fields $\emA_a$ on $\scri^+$, satisfying $\emA_a n^a=0$, with $\emA$
tending to zero at $\spi$.  Again, $\emA_a$ has two independent components
which capture the two radiative modes of electro-magnetic field.  The
symplectic structure can be written as:
\begin{equation}
   \begin{eqtableau}{1}\label{SS}
      \Omega^\Rad (\delta_1,\delta_2) :&=
        \frac{1}{32\pi G} \int_{\scri^+}
\hat{q}^{ac}\hat{q}^{bd}
          \left[ \delta_1 \gamma_{ab} \,
\Lie_{\hat{n}} (\delta_2 \gamma_{cd}) -
            \delta_2 \gamma_{ab} \,
\Lie_{\hat{n}} (\delta_1 \gamma_{cd}) \right]
  \tensor^\scri<\hat{\epsilon}> \\
  &\qquad + \frac{1}{8\pi} \int_{\scri^+} \hat{q}^{ab}
     \left[ \delta_1 \emA_a \, \Lie_{\hat{n}} (\delta_2 \emA_b) -
       \delta_2 \emA_a \, \Lie_{\hat{n}} (\delta_1 \emA_b) \right]
       \tensor^\scri<\hat{\epsilon}> \\
   \end{eqtableau}
\end{equation}
(For details, see Appendix C.1 and \cite{jmp,as}.)

The asymptotic symmetry group at $\scri^+$ is the BMS group \cite{bms}
which admits a preferred four-dimensional Abelian sub-group of
translations.  Let us suppose that the conformal factor is chosen such
that $\hat{q}_{ab}$ is the unit 2-sphere metric.  Then, $\hat{n}^a$ is
a BMS time-translation.  Diffeomorphisms generated by $\hat{n}^a$
induce motions on the radiative phase-space.  As one might expect,
they preserve $\Omega^\Rad$ and the corresponding Hamiltonian is given
by \cite{as}:
\begin{equation}\label{HamScri}
    H^\Rad_{\hat{n}} =
  -\frac{1}{32 \pi G}\int_{\scri^+} N_{ab}N_{cd}\hat{q}^{ac}
    \hat{q}^{bd}\tensor^\scri<\hat{\epsilon}>
  -\frac{1}{8\pi} \int_{\scri^+}
  (\emF_{ac} \tensor<\emF_b^c> + 
  \dual\emF_{ac} \tensor<\dual\emF_b^c>)
  \hat{n}^a \hat{n}^b \tensor^\scri<\hat{\epsilon}> \\
\end{equation}
where $N_{ab} = -2\Lie_{\hat{n}} \gamma_{ab}$ is the Bondi News tensor
at the point in the radiative phase space labeled by $\gamma_{ab}$. 
Thus, $\delta H^\Rad = \Omega^\Rad (\delta, X_{\hat{n}})$ for any
tangent vector $\delta$ to the radiative phase space.  Using the
asymptotic form of the space-time metric in suitable coordinates,
Bondi and Sachs \cite{bms} identified the right side of
(\ref{HamScri}) as
\begin{equation} \label{bms}
     P^\Rad \cdot t  = - E^\Rad
 \end{equation}
where $P^\Rad$ is the 4-momentum radiated across $\scri^+$ and $t$
represents the BMS time translation defined by $\hat{n}^a$.  Thus,
$E^\Rad$ is the flux of energy across $\scri^+$ carried by
gravitational and electro-magnetic waves.  (Again, the negative sign
arises in (\ref{bms}) because our signature is $-+++$.)  The
Hamiltonian formulation at null infinity \cite{as} provides a general
conceptual setting in support of this interpretation.

We wish to relate these structures on the radiative phase space with
those on the canonical phase space introduced in Section \ref{s6}. 
Fix a point on the constraint surface of the canonical phase space and
evolve it using field equations.  Consider tangent vectors satisfying
linearized constraints at this point and evolve them using linearized
field equations.  In appendix C, assuming the background and
linearized solutions satisfy certain falloff conditions, we find
\begin{equation}\label{SymEq}
  \Omega(\delta_1,\delta_2)=\Omega^\Rad
  (\delta^\Rad_1,\delta^\Rad_2).
\end{equation}
where $\Omega^\Rad$ is evaluated at the point in the radiative phase
space defined by the background solution and $\delta^\Rad$ is the
tangent vector in the radiative phase space defined by the linearized
solution $\delta$.  The idea is to let $\delta_1$ be arbitrary, choose
for $\delta_2$ the Hamiltonian vector field defined by a time
translation, and use (\ref{SymEq}) to relate the canonical Hamiltonian
(\ref{Ham}) to the radiative Hamiltonian (\ref{HamScri}).

Let us choose a vector field $t^a$ on $\man$ such that: i)$t^a =
\ell^a$ on $\Delta$; ii) $t^a$ is unit at spatial infinity and defines
an asymptotic time-translation; and, iii) $t^a$ is a BMS time
translation at $\scri^+$ and the conformal factor is so chosen that
$t^a = \hat{n}^a$ at $\scri^+$.  As in Section \ref{s6}, leaves $M$ of
our foliation will be assumed to be asymptotically orthogonal to
$t^a$.  Then, from Section \ref{s6} we have $\delta{H}_t
=\Omega(\delta, X_t)$, with $X_t \equiv (\Lie_{t} A, \Lie_{t}\Sigma;\,
\Lie_{t} {\emA}, \Lie_{t}\emE)$.  From the above discussion of the
radiative phase space, we have: $\delta^\Rad \,{H}^\Rad_t =
\Omega^\Rad(\delta^\Rad, X_t^\Rad)$ where $X^\Rad_t \equiv
(\Lie_{\hat{n}} \gamma;\, \Lie_{\hat{n}}\emA )$.  Therefore, using the
equality (\ref{SymEq}) of the two symplectic structures, we conclude:
\begin{equation}\label{HeqH}
  \delta H_t = \delta^\Rad\, H^\Rad_t=
  -\delta^\Rad\, \ERad
\end{equation}
for all linearized solutions $\delta$ satisfying the asymptotic conditions. 
Let us assume such solutions span the tangent space at every point in the
portion of the phase space under consideration.  Then, we conclude that $H_t$
and $H_t^\Rad$ differ by a constant.  To fix the value of that constant, let
us examine the Reissner--Nordstr\"om space-times.  We already saw in Section
\ref{s7.2} that in these space-times $H_t$ vanishes.  Since they are static,
the Bondi News tensor and the electro-magnetic radiation vanish on $\scri^+$. 
Hence, $H_t^\Rad$ also vanishes in a Reissner--Nordstr\"om solution.  Thus, the
value of the constant is zero and
\begin{equation}\label{HeqERad}
  H_t = -\ERad.
\end{equation}
As one might have intuitively expected, the canonical Hamiltonian
equals the time component of the flux of the 4-momentum that is radiated
across $\scri^+$.

Recall from (\ref{HamVal}) that, when the constraints are satisfied, the value
of the canonical Hamiltonian $H_t$ is equal to the isolated horizon mass
$M_{\Delta}$ minus the ADM energy $E^\ADM$ in the rest frame of the isolated
horizon.  Therefore, it now follows that $M_\Delta = E^\ADM - \ERad$.  It is
well-known that the difference of the ADM energy and the flux of energy
through $\scri^+$, denoted $E^{\rm Bondi}(i^+)$, is the future limit of the
Bondi energy (in the rest frame defined by $t^a$) \cite{amprl}.  Hence, we
conclude:
\begin{equation}\label{meaning}
  M_\Delta=  E^{\rm Bondi}(i^+)
\end{equation}
Thus, $M_\Delta$ can be thought of as the mass remaining in the space-time
after all radiation has escaped to infinity, or, equivalently, the mass of
the black hole with its static hair.  This simple interpretation provides
additional support for our definition of the horizon mass.

As emphasized earlier, the discussion of this subsection is based on a number
of technical assumptions (which are stated in Appendix C.2).  We will conclude
with a summary of their physical content.  Apart from asymptotic flatness at
spatial and null infinity, the key assumptions involve the structure of $i^+$. 
We assume there is only one bound state in asymptotic future, represented by
the isolated horizon.  This reflects the expectation that, in the
Einstein--Maxwell theory, there would be no gravitational or electro-magnetic
radiation hovering around the horizon at late times.  Multiple black hole
solutions which reach equilibrium asymptotically are excluded, as their
structure at $i^+$ would be quite different from that of the Schwarzschild
space-time.  If the black holes do not reach asymptotic equilibrium but
accelerate away from each other, the structure at $i^+$ may be similar to (or
even simpler than) that in the Schwarzschild space-time.  An example is
provided by the C-metric, where, if the parameters are adjusted suitably, the
structure at $i^+$ as well as that at $\spi$ is regular as in Minkowski
space-time.  However, the accelerating black holes pierce $\scri^+$ which is
now singular.  Therefore, as it stands, our analysis is not applicable to this
case either.  Although our analysis could conceivably be generalized to cover
these two types of situation, its current form is primarily applicable to the
situation depicted in figure 1(a) in which a single gravitational collapse
occurs.

\section{First Law}
\label{s8}

Since we now have well-defined notions of surface gravity $\kappa$, electric
potential $\Phi$ and horizon mass $M_\Delta$, we are ready to examine the
question of whether the first law holds.  In section \ref{s8.1}, we will
consider the \textit{equilibrium version} of the law in which the horizon
observables of two nearby space-times are compared.  This version is closer in
spirit to the treatment of the first law of thermodynamics in which one
compares the values of macroscopic, thermodynamic quantities associated with
two nearby equilibrium configurations, without reference to the process which
causes the transition between them.  In section \ref{s8.2}, we will consider
the \textit{physical process version} of the first law in which one explicitly
considers the process responsible for the transition.  This version will bring
out certain subtleties.

\subsection{Equilibrium version}
\label{s8.1}

Denote by $\IH$ the (infinite-dimensional) space of space-times admitting (one
or more) isolated horizons.  In this section, we will be concerned only with
the structure near isolated horizons.  In particular, we will not have to
refer at all to the boundary conditions at infinity or to the precise nature
of the matter outside the isolated horizon.  We will simply assume the surface
gravity $\kappa$, the potential $\Phi$ and the mass $M_\Delta$ of the isolated
horizon are determined by its intrinsic parameters $\rad$ and $Q_\Delta$ via
(\ref{surfGrav}), (\ref{fixMax}) and (\ref{mass}):
\begin{equation} \label{mkappa}
  \begin{eqtableau}[.1cm]{2}
    M_\Delta &= \frac{\rad}{2G} \left(1 + \frac{GQ_\Delta^2}{{\rad}^2}
    - \Lambda {\rad}^2\right)
    &\hspace{1cm}  \area &= 4\pi r_\Delta^2 \\
    \kappa &= \frac{1}{2\rad} \left( 1 - \frac{GQ_\Delta^2}{{\rad}^2}
    -\Lambda {\rad}^2\right)
    &\hspace{1cm}  \Phi &= \frac{Q_\Delta}{r_\Delta}\, ,
  \end{eqtableau}
\end{equation}
and allow general matter fields in the exterior.  (Recall from Section
\ref{s3} that the notion of isolated horizons is not tied down to
Einstein--Maxwell theory.)  This viewpoint is similar to that normally adopted
for the ADM 4-momentum and angular momentum defined at spatial infinity.
These quantities are first derived from Hamiltonian considerations adapted to
specific matter sources (e.g., Klein Gordon, Maxwell, Dirac and Yang-Mills
fields) but then used also for total 4-momentum and angular momentum for more
general forms of matter (e.g. fluids) for which a satisfactory initial value
formulation and Hamiltonian framework may not exist.  Thus, one often uses the
expressions of 4-momentum and angular momentum at infinity without specifying
the precise matter content, assuming only that the stress-energy tensor
satisfies physically reasonable conditions and falls off appropriately.  In
the same spirit, we now assume that $M_\Delta$, $\kappa$ and $\Phi$ of an
isolated horizon $\Delta$ of radius $\rad$ and charge $Q_\Delta$ are given by
(\ref{mkappa}), irrespective of the matter content outside, so long as that
matter does not endow the horizon with additional intrinsic parameters (such
as a dilatonic charge or a new $U(1)$ charge).

Given a space-time $(\man, g_{ab})$ in $\IH$ and a tangent vector
$\delta$ at this point, we can consider a smooth curve in $\IH$
passing through this point with $\delta$ as a tangent vector there and
examine how $M_\Delta$ and $\kappa$ associated with the isolated
horizon $\Delta$ in the background change. Straightforward algebra
yields:
\begin{equation} \label{firstlaw}
\delta M_{\Delta} = \frac{1}{8\pi G} \kappa\delta\area +
\Phi \delta Q_\Delta.
\end{equation}
This equation tells us the relation between infinitesimal changes in
the mass, area, and charge of two nearby, non-rotating isolated
horizons. It is our generalization to isolated horizons of the
{\it equilibrium version} of the first law of black hole mechanics.

We will conclude this sub-section with a few remarks.

\rem{1} The above calculation leading to the first law is trivial.  The
non-trivial part of the analysis was to arrive at expressions (\ref{mkappa})
of $\kappa$ and $M_\Delta$ in absence of a static Killing field.  Again,
although our boundary conditions allow the presence of radiation arbitrarily
close to the horizon, they successfully extract the structure from event
horizons of static black holes that is relevant for thermodynamic
considerations.  As with the zeroth law, the veracity of the first law can be
taken as additional support for our definitions of $\kappa$ and $M_\Delta$.

\rem{2} The laws of black hole mechanics were first derived by Bardeen,
Carter and Hawking \cite{2,bc}.  They considered stationary black
holes possibly surrounded by a perfect fluid in a circular flow and
arrived at the first law by comparing two nearby stationary solutions.
For purposes of comparison, it is more convenient to use an extension
of that work to more general matter fields discussed by Heusler
\cite{heus}.  In the non-rotating case, their main results can be
summarized as follows.  Identities governing the Komar integral of the
static Killing vector imply the mass $M_\ADM$ measured at
spatial infinity is given by
\begin{equation}
M_\ADM = \frac{\kappa}{4\pi G} a_h +\int_M (2T_{ab} - Tg_{ab})\,
K^a\, dS^b
\end{equation}
where $a_h$ is the area of the horizon, $M$ is a partial Cauchy surface from
the horizon to spatial infinity, $T_{ab}$ is the stress-energy of the matter
field, and $K^a$ is the static Killing field.  Under variations $\delta$ from
one such stationary black hole solution to another, it was then shown that
\begin{equation} \label{1law}
\delta{M}^\ADM = \frac{\kappa}{8\pi G}\, \delta A +
\frac{1}{16\pi}\,\int_M G^{ab} (\delta{g}_{ab})\, \dual K -
\frac{1}{8\pi}\, \int_M \delta \dual(G.K)\, ,
\end {equation}
where $\dual K$ is the 3-form dual to $K^a$ and $\dual(G.K)$, the 3-form dual
to $G^{ab}K_b$.  In the Einstein--Maxwell case, it turns out that the two
\textit{volume} integrals in (\ref{1law}) collapse to a single term at the
horizon which is precisely $\Phi\delta Q_\Delta$, where $\Phi \hateq \emA.K$.
Thus, in this case, the first law relates the change in the ADM mass to
changes in quantities at the horizon.  The final form is the same as
(\ref{firstlaw}) (with $M_\Delta$ replaced by $M_\ADM$).  Similarly the final
expression of $M_\ADM$ is just the right side of (\ref{mass}).

However, there are some important differences from our approach.  First, in
the above derivation, one restricts oneself to \textit{static} solutions of
field equations and makes a heavy use of the Komar integral associated with
the Killing vector $K^a$.  Second, the permissible variations $\delta$ are
only those which relate nearby static solutions.  Third, the term $\Phi \delta
Q_\Delta$ at the horizon arose from the volume terms in (\ref{1law}); that its
value depends only on the fields evaluated at the horizon is not fundamental
to this derivation.  The whole calculation of \cite{heus} is based on an
interplay between infinity and the horizon which is possible only because the
assumption of staticity makes the problem elliptic and field variations
`rigid'.  The black hole is not studied as an isolated, separate entity; the
quantities defined at the horizon are tightly tied to the exterior fields. 
Indeed, there is no useful analog of the mass $M_\Delta$, associated with the
horizon.  Perhaps the closest analog is the Komar integral evaluated
\textit{at} the horizon, sometimes denoted $M_H$ \cite{bc}.  However, in the
charged case, this integral does not include the Maxwell contribution $\Phi
Q_\Delta$ and cannot therefore be used directly in the first law.  Our
derivation, by contrast, makes no reference to a Killing vector and allows
radiation fields outside the horizon.  All our considerations are local to the
isolated horizon; in the variation, we did not have to refer to bulk fields on
$M$ at all.  Finally, the mass which appears in our first law is the isolated
horizon mass, $M_\Delta$, and \textit{not} the ADM mass $M_\ADM$.

\rem{3} A treatment of the first law based on Hamiltonian
considerations was given by Wald and collaborators
\cite{sw,rw,rwbook}.  The final result of this work is more general
than that of Bardeen-Carter-Hawking type of analyses.  The background
space-time is again a stationary black hole, possibly with matter
fields in the exterior.  However, the perturbations $\delta$ are no
longer required to be stationary; they can relate the background
stationary solution to any nearby solution.  Since our approach is
also based on the Hamiltonian framework, the two treatments share a
number of common features.

However, there are also a number of important differences, both in methodology
and final results.  While boundary conditions play a key role in our analysis,
their analogs are not specified in references \cite{sw,rw,rwbook}.
Consequently, the issue of differentiability of action is not discussed.  In
particular, while our action $S$ of (\ref{action}) \textit{cannot} be written
as a pure bulk term, in \cite{sw,rw,rwbook} there is no surface term in the
action either at infinity or at the horizon.  Consequently, the Hamiltonian
contains only bulk terms and there is no analog of our horizon mass
$M_\Delta$.  When restricted to non-rotating black holes, the first law of
\cite{sw,rw,rwbook} has the same form as (\ref{firstlaw}).  However, as in the
Bardeen-Carter-Hawking approach, our $M_\Delta$ is replaced by the ADM mass
$M_\ADM$ and the background space-time is assumed to be a stationary black
hole solution.

\rem{4} As indicated in Section \ref{s3.2}, the space $\IH$ of space-times
admitting isolated horizons in Einstein--Maxwell theory is
infinite-dimensional.  The space $\S$ of static solutions is a
finite-dimensional subspace of $\IH$.  In the Bardeen-Carter-Hawking type
approach, the first law holds only at points of $\S$ and for tangent vectors
$\delta$ to $\S$.  In the Wald approach, it holds again only at points of $\S$
but the variations $\delta$ need not be tangential to $\S$.  In the approach
developed in this paper, the first law holds at \textit{all} points of $\IH$
and for \textit{all} tangent vectors to $\IH$.  Thus, the generalization
involved is very significant.  However, since our boundary conditions imply
the intrinsic geometry of $S_\Delta$ is spherically symmetric, distorted black
holes are excluded from our analysis.  By contrast, the other two approaches
\textit{can} handle static, distorted black hole solutions.  In
Einstein--Maxwell theory, there are no such solutions.  However, if we allow
charged fluid sources, such solutions presumably exist.  Therefore, in the
general context, our framework misses out certain situations which are
encompassed by the other two approaches.  It would be interesting to
generalize our framework to overcome this limitation.

\subsection{Physical process version}
\label{s8.2}

Let us now consider the situation depicted in figure 1(b).  We are given a
space-time with a non-rotating, isolated horizon $\Delta_1$ (with parameters
$r_1$ and $Q_1$).  Suppose a small amount of matter falls in to the horizon
and after a brief dynamical period the horizon settles down to a new
equilibrium configuration $\Delta_2$ (with parameters $r_2$ and $Q_2$).  The
question is: How do the observables associated with the horizons change in
this physical process?  The difference from the situation considered in the
previous subsection is that one is now considering a physical process
occurring in a single space-time rather than comparing two nearby space-times.

It is completely straightforward to analyze the process in our framework since
the two masses and surface gravities are determined by their intrinsic
parameters via (\ref{mkappa}).  The actual algebraic calculation is the same
as in the last sub-section; only the physical meaning of the variation
$\delta$ is different.  Hence, we again find the changes in mass, area and
charge are governed by (\ref{firstlaw}).

It is instructive to analyze this relation in terms of properties of the
matter which fell across the horizon.  Using the Raychaudhuri equation, and
keeping only first order terms in variations, it is straightforward to show
\cite{rwbook}
\begin{equation} \label{flux}
\kappa\, \delta a_\Delta = 8\pi G\,\, \delta E^{\rm flux} \equiv
8\pi G\, \int_{\mfs{H}} \dual(\delta T\cdot \ell),
\end{equation}
where $\mfs{H}$ is the portion of the horizon (between $\Delta_1$ and
$\Delta_2$) crossed by the matter and $\dual(\delta T\cdot \ell) $ is the
3-form dual to $\delta T^{a}{}_b \ell^b$.  Let us first suppose $Q_2 =Q_1$,
i.e., the charge of the horizon did not change in this physical process.
Then, comparing (\ref{flux}) with (\ref{firstlaw}), we arrive at a simple
physical picture: the change in the mass of the horizon is equal to the total
energy flux $E^{\rm flux}$ across the horizon.  However, it is interesting to
note that, if $Q_2 \not= Q_1$, there is an extra contribution, $\Phi \delta
Q_\Delta$, to $\delta M_\Delta$.  What is the origin of this term?  It arises
due to `book-keeping' in the following sense.  As we observed in Section
\ref{s7}, $M_\Delta$ contains not just the `raw energy of the content of the
horizon', but also the energy of the electro-magnetic hair outside the
horizon.  (Recall that in static solutions, $M_\Delta$ equals the ADM mass
and, more generally, it equals the future limit of the Bondi mass, both of
which include the contribution from energy in the Coulombic electro-magnetic
field outside the horizon.)  Before the physical process began, the charge
$\delta Q_\Delta = Q_2 - Q_1$ is outside the horizon and the energy in its
Coulomb field does not contribute to $M_1$.  At the end of the process,
however, the black hole charge changes by $\delta Q_\Delta$ and the energy in
the corresponding Coulombic field does contribute to $M_2$.  This accounts for
the term $\Phi \delta{Q_\Delta}$ in the expression of $\delta{M}_\Delta = M_2
-M_1$.  Thus, the physical process version of the first law is subtle.  The
first order change in the mass of the horizon has a two-fold origin: a
contribution due to flux of energy across horizon and another contribution
from book-keeping of the energy in the Coulombic hair of the horizon.

What is the situation in the standard framework, where one uses
$M_\ADM$ in place of $M_\Delta$?  To our knowledge, the physical
process version of the first law has been discussed only in the
uncharged case \cite{rwbook}.  One assumes that the background
space-time is globally static and considers a (non-static) matter
perturbation which falls across the horizon. The ADM mass of the
unperturbed space-time is taken to be $M_1$ and the ADM mass of the
background plus perturbation is taken to be $M_2$.  Then, using the
reasoning given above, one arrives at the first law $\delta M =
(\kappa \delta a)/8\pi G$ and interprets $\delta M$ as the change in
the mass of the black hole due to the energy flux across the horizon.
However, in the charged case, if $\delta Q_\Delta$ is not equal to
zero, it seems difficult to account for the term $\Phi \delta
Q_\Delta$ which also contributes to $\delta M$ without bringing in
$M_\Delta$.%
\footnote{At first sight, it may appear that one should be able to
account for this term using Maxwell's equations to simplify the
contribution to $\int_\Delta \, \dual(\delta T \cdot \ell)$ that
comes from the first order change in the stress energy of the Maxwell
field.  However, a closer examination shows that the integral of the
Maxwell contribution vanishes identically.}
It is interesting to note that, in this respect, there is a key difference
between the angular momentum work term $\Omega \, \delta J$ and the
electro-magnetic term $\Phi\, \delta Q_\Delta$: While the angular momentum
contribution is coded easily in the flux of the stress-energy across the
horizon, the electro-magnetic contribution is not.  This is why, unlike the
electro-magnetic work term, the angular momentum work term can be easily
incorporated in the physical process version in the standard approach
\cite{rwbook}.

\section{Discussion}
\label{s9}

Let us begin with a summary of the main ideas and results.

In Section \ref{s3} we introduced the notion of a non-rotating
isolated horizon $\Delta$.  While one needs access to the entire
space-time to locate an event horizon, isolated horizons can be
located quasi-locally.  Event horizons of static black holes in
Einstein--Maxwell theory do qualify as isolated horizons.  However, the
definition does not require the presence of a Killing field even in a
neighborhood of $\Delta$.  Rather, physically motivated, geometric
conditions are imposed on the null normal $\ell^a$ to $\Delta$ and on
an associated inward pointing null vector $n^a$ \textit{at} $\Delta$. 
These conditions imply the Lie derivative along $\ell^a$ of the
intrinsic (degenerate) metric of $\Delta$ vanishes which in turn
implies the area of an isolated horizon is constant in time.  In this
sense, the horizon itself is isolated or `in equilibrium'.  However,
the space-time may well admit electro-magnetic and/or gravitational
radiation.  The quasi-local nature of the definition of $\Delta$ and
the possibility of the presence of radiation suggest the space of
solutions to the Einstein--Maxwell equations admitting isolated
horizons would be \textit{infinite}-dimensional, in striking contrast
to the space of \textit{static} black holes which is only
three-dimensional.  Recent mathematical results by a number of workers
\cite{jl,hf,r,schoen} show this expectation is indeed correct.

While the conditions used in the definition seem mild, they lead to a
surprisingly rich structure.  In particular, the intrinsic metric of
$\Delta$, the shear, twist and expansion of $\ell^a$ and $n^a$ and
several of the Newman-Penrose gravitational and electro-magnetic
curvature scalars \textit{at} $\Delta$ have the same functional
dependence on the radius $\rad$ and charges $Q_\Delta, P_\Delta$ as in
the Reissner--Nordstr\"om solutions.  This rich structure enables one
to fix naturally the scaling of the null normal $\ell^a$ and leads to
an unambiguous definition of surface gravity, $\kappa$.  Furthermore,
using only the structure available at $\Delta$, one can show that
$\kappa$ is constant on $\Delta$; the zeroth law is thus extended from
static black holes to isolated horizons.

To formulate the first law, we need a notion of the mass $M_\Delta$ of the
isolated horizon.  Since we allow for the presence of radiation outside
$\Delta$, we cannot use the ADM mass $M_\ADM$ as $M_\Delta$, nor do we have a
static Killing field to perform a Komar integral at $\Delta$.  Fortunately, we
can use the Hamiltonian framework.  Although the presence of the internal
boundary introduces several subtleties, a satisfactory Hamiltonian framework
\textit{can be} constructed.  When the constraints are satisfied, the
Hamiltonian turns out to be a sum of two surface terms, one at infinity and
one at $\Delta$.  As usual, the term at infinity yields the ADM energy and we
define $M_\Delta$ to be the surface term at $\Delta$.  This definition is
supported by several independent considerations.  In particular, under
suitable conditions, $M_\Delta$ turns out to be the future limit of the Bondi
mass.  Having expressions for both $\kappa$ and $M_\Delta$ at our disposal, we
ask if the first law holds.  The answer is in the affirmative for both the
`equilibrium state' and the `physical process' versions.  This provide a
significant generalization of the first law of mechanics of static black holes
in the Einstein--Maxwell theory.  Furthermore, in the charged case, this
analysis brings out some subtleties associated with the `physical process'
version.  However, since our framework focusses on isolated horizons and small
perturbations thereof, it does not shed new light on the second law of black
hole mechanics which refers to fully dynamical situations.

These underlying ideas overlap with those introduced in references
\cite{9,ack}.  In \cite{9}, Hayward introduced, and very effectively
used, the notion of `trapping horizons'.  Our isolated horizons are a
special case of trapping horizons, the most important restriction
being our assumption that the expansion of the horizon is zero.  This
assumption is essential to capture the notion that the horizon is in
equilibrium, which underlies the zeroth and the first law. 
Furthermore, our method of defining the surface gravity $\kappa$ and
the mass $M_\Delta$ of isolated horizons differ from those used by
Hayward for trapping horizons and consequently our treatment of the
two laws is also different.  (To our knowledge, in the context of
trapping horizons, a satisfactory definition of surface gravity is
available only for spherically symmetric space-times.)  However, the
notion of isolated horizon is clearly inadequate for the treatment of
dynamical situations which are considered, for example, in the second
law and it is these situations that provide a primary motivation in
the analysis of trapping horizons.

The relation between the ideas discussed in this paper and those
introduced in \cite{ack} is closer.  Both papers deal with isolated
horizons.  However, while the focus of reference \cite{ack} is on
constructing a Hamiltonian framework suitable for quantization and
entropy calculations, the focus of the present paper is on the
mechanics of isolated horizons.  The two overlap in their
constructions of Hamiltonian frameworks.  However, as explained in
Section \ref{s6}, reference \cite{ack} only considers isolated
horizons with \textit{fixed} parameters $\rad, Q_\Delta,P_\Delta$ and
therefore ignores several subtleties which are critical to our present
treatment of the first law.  Reciprocally, in \cite{ack}, significant
effort went into the construction of a Hamiltonian framework in terms
of \textit{real} variables which is necessary for quantization but not
for the laws of mechanics.  Finally, in Sections \ref{s3} and \ref{s4}
and in Appendices A and B, we took the opportunity to present the
necessary background material from a perspective which is different
from but complementary to that adopted in \cite{ack}.

We will conclude by indicating a few avenues to extend the present
work.

\rem{1} Let us begin with the non-rotating case.  Although we did not
explicitly require the intrinsic metric of an isolated horizon to be
spherically symmetric, our assumptions on properties of the null
vector fields $\ell^a$ and $n^a$ \textit{at} the horizon led us to
this conclusion.  The discussion of Section \ref{s2.2} shows the
assumptions are not overly restrictive: the
class of space-times satisfying them is infinite-dimensional.%
\footnote{This may seem surprising at first since most of the current
intuition comes from static black holes and, in the static context,
generalizations of Reissner--Nordstr\"om solutions naturally lead to distorted
horizons.  However, this is because static problems are governed by elliptic
equations and generic perturbations in the exterior then force the horizon
itself to be distorted.  Radiative space-times provide generalizations in
quite a different direction.  Now, the equations are hyperbolic and, as the
`gluing methods' of Corvino and Schoen \cite{schoen} show, the geometry can be
spherical even in a neighborhood of the horizon without being spherical
everywhere.  More generally, as the Robinson--Trautman solutions illustrate
\cite{pc}, the rotational Killing fields may not extend even to a neighborhood
of the horizon.}
Typically, these space-times will admit radiation and will \textit{not} be
spherically symmetric in the bulk.  Nonetheless, it is of interest to weaken
our assumptions to allow space-times with `distorted' horizons on which the
intrinsic geometry will not be spherical.  For simplicity, consider the case
in which there is no matter in a small neighborhood of $\Delta$.  Then, we
would expect only to have to weaken the conditions on $n^a$ and allow $\mu$ to
be non-spherical.  The structure at event horizons of static, distorted
black-holes has been recently examined by Fairhurst and Krishnan and their
analysis confirms this hypothesis.  The extension of the framework presented
here to incorporate distortion should be fairly straight forward.

\rem{2} Inclusion of rotation would provide an even more interesting
extension.  Again, conditions on $\ell^a$ would remain unchanged.  Only the
conditions on $n^a$ and the sphericity requirement on the component
$T_{ab}\ell^a n^b$ of stress-energy tensor at $\Delta$ will have to be
weakened.  In particular the Newman-Penrose spin-coefficient $\pi$ can no
longer be zero since it is a potential for the imaginary part of $\Psi_{2}$
which carries the angular momentum information. Work is already in progress
on this generalization.

\rem{3} In the stationary context, using Hamiltonian methods and
Noether charges, Wald \cite{rw} has extended the notion of entropy and
discussed the first law in a wide variety of gravitational theories,
possibly coupled to bosonic fields, in any space-time dimension.  It
would be very interesting to extend the present framework for isolated
horizons in a similar fashion.  As a first step, one would recast the
framework in terms of tetrads $e^a_I$ and the associated real, Lorentz
connections $A_{aI}{}^{J}$.  The extension of the resulting
(Einstein-matter) action and Hamiltonian framework to higher
dimensions should then be straight-forward.  The first step is easy to
carry out since the tetrads can be easily obtained from soldering
forms and the Lorentz connection is just the real part of our
self-dual connection.  Thus, the 4-dimensional tetrad action is, in
effect, just the real part of (\ref{action}) and the corresponding
Hamiltonian is just the real part of (\ref{Ham}).  Hence, it should be
rather easy to extend the present results to higher-dimensional
general relativity, possibly coupled to matter.  Furthermore, since
the basic variables are tetrads rather than metrics, it should be
straightforward to allow fermionic matter as well.  Incorporating
general gravitational theories, on the other hand, could be highly
non-trivial.

\bigskip\bigskip

{\bf {Acknowledgments}} We are most grateful to John Baez, Alejandro
Corichi, Kirill Krasnov and Jerzy Lewandowski for many stimulating
discussions.  We have also profited from comments made by numerous
colleagues, especially Brandon Carter, Piotr Chru\'sciel, Helmut
Friedrich, Sean Hayward, Don Marolf, Istvan Racz, Oscar Reula, Carlo
Rovelli, Bernd Schmidt, Thomas Thiemann and Robert Wald.  The authors
were supported in part by the NSF grants PHY94-07194, PHY95-14240,
INT97-22514 and by the Eberly research funds of Penn State.

\appendix

\section{Conventions}
\label{Conv}

In this paper, capital primed and unprimed indices represent $\rm
SL(2,\Com)$ spinors fields.  As usual, spinors with only unprimed
indices are also interpreted as $\rm SU(2)$ spinors in the phase space
framework.  The spinor conventions are largely those of \cite{pr1},
but with minor modifications to replace the $+---$ signature of
\cite{pr1} with the $-+++$ signature used here.  We describe these
modifications here.

\subsection{Metric and Null Tetrad}

The metric is given in terms of the soldering form by
\begin{equation}
  g_{ab} = \tensor <\sigma_a^AA'> <\sigma_b_AA'>.
\end{equation}
Because of our choice of signature, however, the soldering form must
be taken to be \textit{anti}-Hermitian: $\tensor <\bar\sigma_a^AA'> `=
-' <\sigma_a^AA'>$.

A spin dyad $(\i^A, \o^A)$, satisfying $\i^A \o_A=1$, defines a null tetrad as
follows:
\begin{equation}
  \begin{eqtableau}{2}
    \ell^a   &= i\tensor<\sigma^a_AA'><\o^A><\bar \o^A'> &\qquad
    m^a      &= i\tensor<\sigma^a_AA'><\o^A><\bar \i^A'> \\
    n^a      &= i\tensor<\sigma^a_AA'><\i^A><\bar \i^A'> &\qquad
    \bar m^a &= i\tensor<\sigma^a_AA'><\i^A><\bar \o^A'>. \\
  \end{eqtableau}
\end{equation}
This tetrad obeys the usual inner product conventions in the $-+++$
signature: the only non-vanishing inner products are $\ell^a n_a = -1$
and $m^a \bar m_a = 1$. (Note that the definitions of $m$ and
$\bar{m}$ differ from those used in \cite{ack}. This change was
necessary to make the values of our Newman-Penrose curvature
components the same as those found in the literature even though our
signature is $-+++$.)

\subsection{Volume Forms and Orientations}

The volume form on space-time is defined by its spinor expression, which is the
same as that used in \cite{pr1}:
\begin{equation}
  \tensor ^4<\epsilon_abcd> `=' <\sigma_a^AA'> <\sigma_b^BB'>
    <\sigma_c^CC'> <\sigma_d^DD'> `\left[ -i' <\epsilon_AB> <\epsilon_CD>
      <\epsilon_A'C'> <\epsilon_B'D'> + \mathrm{c.c.} \right].
\end{equation}
This volume form can be expressed in terms of the null tetrad as
\begin{equation}
  \tensor ^4<\epsilon_abcd> `= 24i\,' <\ell_[a> <n_b> <m_c> <\bar m_d]>.
\end{equation}

The conventions for inducing volume forms on sub-manifolds of space-time
are designed to be compatible with those used in \cite{blue} and with
the usual orientation conventions used in Stokes' theorem on
Riemannian manifolds.  Specifically, this means that a volume form is
induced on a space-like sub-manifold of space-time by contracting its
future-directed, unit normal with the \textit{last} index of $\tensor
^4<\epsilon>$.  Then, within a space-like hypersurface, a volume form
is induced on a two-dimensional sub-manifold by contracting its
outward-bound, unit normal with the \textit{first} index of the volume
form on the hypersurface.  All other orientation conventions can be
determined from these two.  In particular,
\begin{eqset}
\tensor ^3<\epsilon_abc> &= \tensor ^4<\epsilon_abcd> <\hat\tau^d> \\
\tensor ^2<\epsilon_bc> &= \tensor <\hat r^a\:_{\rm in}> ^3<\epsilon_abc>
   `= 2i\,' <m_[b> <\bar m_c]> \\
\tensor ^\Delta<\epsilon_abc> &= -3\tensor ^2<\epsilon_[ab> <n_c]>
   `= -6i\,' <n_[a> <m_b> <\bar m_c]>. \\
\end{eqset}
Here, $\tensor ^3<\epsilon>$ denotes the induced volume form on a
space-like hypersurface, $\tensor ^2<\epsilon>$ denotes the volume form
on one of the $S_\Delta$, and $\tensor ^\Delta<\epsilon>$ denotes the
preferred alternating tensor on the null surface $\Delta$.  Meanwhile,
$\hat\tau$ denotes the future-directed future normal to the space-like
hypersurface and $\hat r^a_{\rm in}$ denotes the unit radial vector
directed \textit{inward} at the horizon. Note that the inward normal
is appropriate because $S_\Delta$ is the inner boundary of the
space-like surface $M$.

\subsection{Self-Dual Basis}

The soldering form defines a basis of self-dual 2-forms on
space-time via
\begin{equation}
  \tensor <\Sigma_ab^AB> `= 2' <\sigma_[a^AA'> <\sigma_b]^B_A'>
    `= 2' <\sigma_a^A'(A> <\sigma_b^B)_A'>.
\end{equation}
Using the spin dyad, these self-dual 2-forms can be expressed as
\begin{equation}\label{SigComp}
\tensor <\Sigma_ab^AB> = 4 \, \ell_{[a} m_{b]} \, \i^A \i^B
+ 4 \, (m_{[a} \bar m_{b]} - \ell_{[a} n_{b]}) \, \i^{(A} \o^{B)}
- 4 \, n_{[a} \bar m_{b]} \, \o^A \o^B.
\end{equation}
One can check that these are, in fact, self-dual in that $\dual\tensor
<\Sigma^AB> `= i' <\Sigma^AB>$.

\subsection{Newman-Penrose Components}

We define the Riemann curvature tensor to be
\begin{equation}
    \tensor<R_abc^d>k_d = 2\nabla_{[a}\nabla_{b]}k_c.
\end{equation}
    The space-time curvature spinors are defined by decomposing the Riemann
tensor as
\begin{eqset}\nonumber
\tensor <R_abcd> &= \tensor <\sigma_a^AA'> <\sigma_b^BB'>
<\sigma_c^CC'> <\sigma_d^DD'> \\ \label{RiemSpin} &\quad
\times \left\{ \epsilon_{A'B'} \epsilon_{CD} \Phi_{ABC'D'} +
\epsilon_{A'B'} \epsilon_{C'D'} \left[ \Psi_{ABCD} - \frac{1}{12} R
\epsilon_{(A(C} \epsilon_{D)B)} \right] + \mathrm{c.c.} \right\}, \\
\interject{leading to the expression for the Ricci tensor:}
\tensor <R_ab> ={}& \tensor <\sigma_a^AA'> <\sigma_b^BB'> \left\{
-2 \Phi_{ABA'B'} + \frac{1}{4} R \epsilon_{AB} \epsilon_{A'B'} \right\}. \\
\end{eqset}
Since the Ricci tensor is real, $\Phi_{ABA'B'}$ is Hermitian.

In terms of the curvature spinors, we define the Newman-Penrose components of
the Weyl tensor $C_{abcd}$ by
\begin{equation}\label{NPPsi}
  \begin{eqtableau}{2}
    \Psi_0 &= \Psi_{ABCD}\o^A\o^B\o^C\o^D
           &=  \hfill&C_{abcd} \ell^a   m^b \ell^c   m^d \\
    \Psi_1 &= \Psi_{ABCD}\o^A\o^B\o^C\i^D
           &= \hfill  &C_{abcd} \ell^a   m^b \ell^c   n^d \\
    \Psi_2 &= \Psi_{ABCD}\o^A\o^B\i^C\i^D
           &=    \hfill&C_{abcd} \ell^a   m^b \bar m^c n^d \\
    \Psi_3 &= \Psi_{ABCD}\o^A\i^B\i^C\i^D
           &=  \hfill&C_{abcd} \ell^a   n^b \bar m^c n^d \\
    \Psi_4 &= \Psi_{ABCD}\i^A\i^B\i^C\i^D
           &= \hfill&C_{abcd} \bar m^a n^b \bar m^c n^d. \\
  \end{eqtableau}
\end{equation}
Note that these definitions are the same as those found in the literature
\cite{pr1,chandra} and, despite the difference in signature, the functions
$\Psi_n$ take their usual values in specific space-times.  Similarly, the
expressions for the Newman-Penrose components of the Ricci tensor read {\small
\begin{equation}\label{NPPhi}
  \begin{eqtableau}{0}
    \begin{eqtableau}{2}
      \Phi_{00} &= \Phi_{ABA'B'} \o^A \o^B \bar\o^{A'} \bar\o^{B'}
               &&= \half R_{ab} \ell^a   \ell^b   \\
      \Phi_{01} &= \Phi_{ABA'B'} \o^A \o^B \bar\o^{A'} \bar\i^{B'}
               &&= \half R_{ab} \ell^a   m^b      \\
      \Phi_{02} &= \Phi_{ABA'B'} \o^A \o^B \bar\i^{A'} \bar\i^{B'}
               &&= \half R_{ab} m^a      m^b      \\
      \Phi_{10} &= \Phi_{ABA'B'} \o^A \i^B \bar\o^{A'} \bar\o^{B'}
               &&= \half R_{ab} \ell^a   \bar m^b \\
    \end{eqtableau}\qquad
    \begin{eqtableau}{2}
      \Phi_{22} &= \Phi_{ABA'B'} \i^A \i^B \bar\i^{A'} \bar\i^{B'}
               &&= \half R_{ab} n^a      n^b      \\
      \Phi_{21} &= \Phi_{ABA'B'} \i^A \i^B \bar\o^{A'} \bar\i^{B'}
               &&= \half R_{ab} \bar m^a n^b      \\
      \Phi_{20} &= \Phi_{ABA'B'} \i^A \i^B \bar\o^{A'} \bar\o^{B'}
               &&= \half R_{ab} \bar m^a \bar m^b \\
      \Phi_{12} &= \Phi_{ABA'B'} \o^A \i^B \bar\i^{A'} \bar\i^{B'}
               &&= \half R_{ab} m^a      n^b      \\
    \end{eqtableau}\\
      \Phi_{11}  = \Phi_{ABA'B'} \o^A \i^B \bar\o^{A'} \bar\i^{B'}
                 = \tsfrac{1}{4} R_{ab} (\ell^a n^b + m^a \bar m^b). \\
  \end{eqtableau}
\end{equation}}%
As before, these are the standard spinorial definitions for the $\Phi_{ij}$.
However, their expressions in terms of the Ricci tensor differ from those of
\cite{pr1} by a minus sign.  This difference occurs because our Ricci tensor
is the negative of the one used in \cite{pr1}.

A similar decomposition can be performed on the electro-magnetic field.  The
Maxwell spinor is defined by expressing the field strength as
\begin{equation}
  \emF_{ab} = \tensor <\sigma_a^AA'> <\sigma_b^BB'>
      (\phi_{AB} \epsilon_{A'B'} + \epsilon_{AB} \bar\phi_{A'B'})
\end{equation}
Then, the Newman-Penrose components of the Maxwell field are defined by
\begin{equation}\label{NPphi}
  \begin{eqtableau}{1}
    \phi_0 &= \phi_{AB} \o^A \o^B = - \ell^a m^b \emF_{ab} \\
    \phi_1 &= \phi_{AB} \i^A \o^B = -\half (\ell^a n^b - m^a \bar m^b)
\emF_{ab}
      = \half m^a \bar m^b (\emF - i \dual \emF)_{ab} \\
    \phi_2 &= \phi_{AB} \i^A \i^B =  n^a \bar m^b \emF_{ab}. \\
  \end{eqtableau}
\end{equation}
As with the gravitational field, the values of these functions will be the
same as those found in the literature.

Finally, the Newman-Penrose spin-coefficients used in this paper are
given by:
\begin{equation}
\begin{eqtableau}{3}
\mu &= m^a \bar{m}^b \nabla_a n_b, &\quad
\lambda &= \bar{m}^a\bar{m}^b\nabla_an_b, &\quad
\pi &= \ell^a\bar{m}^b \nabla_an_b, \\
\sigma &= -m^a m^b\nabla_a\ell_b, &\quad
\rho &= -\bar{m}^a m^b \nabla_a \ell_b, &\quad
\epsilon +\bar{\epsilon} &= -\ell^a n^b \nabla_a \ell_b.
 \\
\end{eqtableau}
\end{equation}
Note that, as is common in the black-hole literature, we denote the
surface gravity by $\kappa$ (so that our $\kappa$ equals $(\epsilon
+\bar\epsilon)$ in the Newman-Penrose notation.) We never need to
refer to the Newman-Penrose spin coefficient $\kappa$.

\section{Newman-Penrose Components and Self-Dual Curvature}
\label{curvApp}

The purpose of this appendix is to establish the relation between the
self-dual ${\rm SL}(2,\Com)$ curvature used in \cite{blue} and the
Newman-Penrose curvature components described in \cite{pr1}.

In any putative space-time where only the Gauss law (\ref{stGauss}) is solved,
the self-dual curvature $F^{AB}$ is equal to the self-dual portion of the
Riemann curvature defined by
\begin{equation}
  \tensor ^+<R_ab^AB> = \half \tensor <\sigma_a^CC'> <\sigma_b^DD'>
    <R_CC'DD'^AA'BB'> <\bar\epsilon_A'B'>
\end{equation}
If we now substitute for the Riemann spinor in this expression using
(\ref{RiemSpin}), one can rearrange the terms to yield
\begin{equation}
\tensor ^+<R_ab^AB> `= -\half' <\bar\Sigma_ab^A'B'> <\Phi^AB_A'B'>
`- \half' <\Sigma_ab^CD> <\Psi^AB_CD> `- \frac{R}{24}' <\Sigma_ab^AB>.
\end{equation}
It is now a long, but straightforward, process to break this formula
into spinor components using (\ref{SigComp}).  Then, using the
definitions (\ref{NPPsi}) and (\ref{NPPhi}) of the Newman-Penrose
components, one can express the result as
\begin{equation}\label{NPCurv}
  \begin{eqtableau}{1}  \tensor ^+<R_{AB}>
           =& \left[ \vphantom{\frac{R}{24}}
               (\Psi_3 + \Phi_{21}) \ell
\wedge n -
                \Psi_4 \ell \wedge m -
                \Phi_{22} \ell \wedge \bar m + {} \right. \\
      &\qquad \left. \vphantom{\frac{R}{24}}
                \Phi_{20} n \wedge m +
                \left( \Psi_2 + \frac{R}{12} \right) n \wedge \bar m -
                (\Psi_3 - \Phi_{21}) m \wedge \bar m \right] o_A o_B \\
           &- \left[ \vphantom{\frac{R}{24}}
                 \left( \Psi_2 + \Phi_{11} -
\frac{R}{24} \right) \ell \wedge n -
                 \Psi_3 \ell \wedge m -
                 \Phi_{12} \ell \wedge \bar m + {} \right. \\
      &\qquad \left. \vphantom{\frac{R}{24}}
                 \Phi_{10} n \wedge m +
                 \Psi_1 n \wedge \bar m -
                 \left( \Psi_2 - \Phi_{11} - \frac{R}{24} \right)
                   m \wedge
\bar m \right] 2 \iota_{(A} o_{B)} \\
           &+ \left[ \vphantom{\frac{R}{24}}
                 (\Psi_1 + \Phi_{01}) \ell
\wedge n -
                 \left( \Psi_2 + \frac{R}{12} \right) \ell \wedge m -
                 \Phi_{02} \ell \wedge \bar m + {} \right. \\
      &\qquad \left. \vphantom{\frac{R}{24}}
                 \Phi_{00} n \wedge m +
                 \Psi_0 n \wedge \bar m -
                 (\Psi_1 - \Phi_{01}) m \wedge \bar m \right] \iota_A
\iota_B.\\
  \end{eqtableau}
\end{equation}
This expresses the self-dual curvature in terms of the Newman-Penrose
components in a spin dyad satisfying $\i^A \o_A = 1$.  The null tetrad here is
defined, of course, by the same dyad.

\section{Symplectic Structure at Null Infinity}

In section \ref{s6} we used the Legendre transform to introduce a symplectic
structure (\ref{Sym}) on the canonical phase space.  On the other hand, there
is also a natural symplectic structure on the space of radiative modes of the
Einstein--Maxwell system, defined intrinsically at (future) null infinity
$\scri^+$.  In this appendix, using field equations, we will show the two
symplectic structures are equal in an appropriate sense, provided the fields
under consideration have suitable asymptotic behavior.

Throughout this discussion, we will restrict ourselves to the region $\man$ of
figure 1(a) which has $\scri^+$ as its future boundary and which admits
partial Cauchy surfaces $M$ which extend from the isolated horizon $\Delta$ to
spatial infinity $\spi$.  As in Section \ref{s7.3}, we will set the
cosmological constant $\Lambda$ to zero.

\subsection{Phase Space of Radiative Modes at $\scri^+$}

Fix an asymptotically flat space-time $(\man,g_{ab})$ and consider its
Penrose completion $(\hat{\man}, \hat{g}_{ab})$. As usual,
$\hat{g}_{ab}=\Omega^2 g_{ab}$ is the conformally rescaled metric and
$\scri^+$ is the future null boundary of $\man$ where the conformal
factor $\Omega$ vanishes. All fields appearing with a `hat' will
refer to the geometry defined by the conformally rescaled metric
$\hat{g}_{ab}$ which is smooth at $\scri^+$.

Let us begin by recalling the `universal structure' at null infinity of
asymptotically flat space-times.  First, $\scri^+$ is topologically $S^2
\times \Re$.  Second, the conformally rescaled metric naturally defines an
intrinsic, degenerate metric $\hat{q}_{ab}= \pback{\hat{g}_{ab}}$ and a null
normal field $\hat{n}_a=\hat{\nabla}_a\Omega$ on $\scri^+$.  We shall assume
the conformal factor $\Omega$ is so chosen as to make $\scri^+$ divergence
free in the sense that $\hat{\nabla}_a \hat{n}^a=0$ on $\scri^+$.  By
construction, $\hat{n}^a$ defines the unique degenerate direction of the
intrinsic metric $q_{ab}$: $\hat{n}^a\hat{q}_{ab}=0$.  Therefore, the `inverse
metric' is unique only up to the addition of a term of the type
$\hat{n}^{(a}\hat{v}^{b)}$ where $\hat{v}^{b}$ is an arbitrary vector field on
$\scri^+$.  (Irrespective of the choice of $v^a$, we have
$\hat{q}_{ab}\hat{q}^{bc}\hat{q}_{cd}=\hat{q}_{ad}$).  Finally, the volume
3-form $\tensor^\scri<\hat{\epsilon}>$ on $\scri^+$ can be defined as
\begin{equation}\label{VolScri}
\tensor^\scri<\hat{\epsilon}_abc>:= \tensor^4<\hat{\epsilon}_abcd>\hat{n}^d
\end{equation}
where $\tensor ^4<\hat\epsilon_abcd>$ is the volume 4-form defined by
the rescaled metric $\hat{g}_{ab}$. These structures are universal in
the sense that they are common to all asymptotically flat space-times;
they do not carry any information about, e.g., the radiation field
which can vary from one space-time to another.

Note however that there remains a conformal freedom in the rescaled metric
$\hat{g}_{ab}$.  If $\Omega$ is an allowable conformal factor which makes
$\scri^+$ divergence-free, so is by $\Omega^\prime = \omega\Omega$, where
$\omega$ is nowhere vanishing on $\scri^+$ and $\Lie_{\hat{n}}\omega=0$ at
$\scri^+$.  Under this transformation, the conformal metric is rescaled as
$\hat{g}_{ab} \rightarrow \omega^2 \hat{g}_{ab}$.  As a consequence, the
pairs, $(\hat{q}_{ab},\hat{n}^a)$ and $(\omega^2 \hat{q}_{ab},
\omega^{-1}\hat{n}^a)$ are to be regarded as (conformally) equivalent at
$\scri^+$.

We can now turn to the dynamical structures and introduce the radiative modes.
Note first that the derivative operator $\hat{\nabla}$ defined by the metric
$\hat{g}_{ab}$ on $\hat{\man}$ naturally induces a derivative operator
$\hat{D}$ defined intrinsically on $\scri^+$ via the pull-back
\begin{equation}
\hat{D}_a\hat{\underline{K}}_b=\pback{\hat{\nabla}_a\hat{K}_b},
\end{equation}
where $\hat{\underline{K}_a}$ is an arbitrary co-vector field defined
intrinsically at $\scri^+$ and $\hat{K}_b$ is any extension of
$\hat{\underline{K}_b}$ to $\man$.  Since $\hat{\nabla}$ is metric compatible,
it follows that $\hat{D}_a \hat{q}_{bc}=0$ and $\hat{D}_a\hat{n}^b=0$.  The
radiative modes of the gravitational field in general relativity are fully
encoded in connections $\hat{D}$ on $\scri^+$ satisfying the above conditions. 
Recall, however, that there is a residual conformal freedom at $\scri^+$.  As
a consequence, one is led to introduce an equivalence relation between
connections.  The phase space of radiative modes consists of these equivalence
classes.  It thus has the structure of an affine space.  The difference
between any two connections in different equivalence classes can be encoded in
a tensor field $\gamma_{ab}$ which satisfies
\begin{equation}
 \begin{eqtableau}{3}\label{GamCond}
\gamma_{ab}\hat{q}^{ab}&=0  &\hspace{1cm} \gamma_{ab} \hat{n}^b&=0
&\hspace{1cm} \gamma_{ab}&=\gamma_{(ab)}.
\end{eqtableau}
\end{equation}
Therefore, by fixing a point in the phase space as the `origin', we
can label any other point by the corresponding tensor field
$\gamma_{ab}$. It is easy to see that $\gamma_{ab}$ has two independent
components which represent the two physical degrees of freedom of
gravitational radiation.

The radiative degrees of freedom of the electro-magnetic field can also
be described by fields intrinsic to $\scri^+$.  It turns out that $\emF_{ab}$
is completely characterized by the unique connection $\emA_a$ at $\scri^+$
satisfying
\begin{equation}\label{emACond}
\begin{eqtableau}{3}
\emA_a \hat{n}^a&=0 &\hspace{1cm} &\mbox{and} &\hspace{1cm} \lim_{u
\rightarrow -\infty}\emA_a&=0
\end{eqtableau}
\end{equation}
where $u$ is the affine parameter along $\hat{n}^a$.  The connection
$\emA_a$ satisfying the above conditions has two independent components.
These represent the two radiative degrees of freedom of the Maxwell
field.

Thus, the phase space of radiative modes at $\scri^+$ consists of pairs
$(\gamma_{ab},\emA_a)$ satisfying the conditions (\ref{GamCond}) and
(\ref{emACond}) respectively.  The symplectic structure on this phase
space is
\begin{equation}\label{SymScri}
\begin{eqtableau}{1}
\Omega^\Rad(\delta^\Rad_1,\delta^\Rad_2):&=
\frac{1}{32\pi G} \int_{\scri^+} \hat{q}^{ac}\hat{q}^{bd}
[\delta_1 \gamma_{ab} \, \Lie_{\hat{n}} (\delta_2 \gamma_{cd}) -
\delta_2 \gamma_{ab} \, \Lie_{\hat{n}} (\delta_1 \gamma_{cd})]
\tensor^\scri <\hat{\epsilon}>
\\
&\qquad +\frac{1}{8\pi} \int_{\scri^+}\, \hat{q}^{ab}
[\delta_1 \emA_{a} \, \Lie_{\hat{n}} (\delta_2 \emA_b) -
\delta_2 \emA_{a} \, \Lie_{\hat{n}} (\delta_1 \emA_b)]
\tensor^\scri <\hat{\epsilon}>.
\end{eqtableau}
\end{equation}

For further details, see \cite{jmp,as}.

\subsection{Equality of Symplectic Structures}

Let us now return to the canonical phase space of Section \ref{s6}.  Fix a
point on the constraint hypersuface and consider tangent vectors which satisfy
the linearized constraints.  Evolve these fields using the appropriate field
equations.  Then, assuming the resulting 4-geometry and the linearized fields
thereon satisfy appropriate falloff conditions, they would provide a point in
the radiative phase space at $\scri^+$ and tangent vectors at that point. 
Using this correspondence, we will now show the canonical symplectic structure
(\ref{Sym}) associated with a partial Cauchy surface $M$ (of figure 1(a))
equals the radiative symplectic structure (\ref{SymScri}) at $\scri^+$.  (The
calculation is modeled after \cite{am2} which discussed the relation between
the two symplectic structures in the absence of internal boundaries within the
framework of geometrodynamics.)  For simplicity of presentation, we will just
make assumptions on the asymptotic behavior of fields as they are needed in
the intermediate stages of the calculation and collect our assumptions at the
end.

The canonical symplectic structure (\ref{Sym}) can be obtained by
integrating a symplectic current $\omega$ on the partial Cauchy
surface $M$. This 3-form $\omega$ is given by
\begin{equation}\label{SymCurr}
\begin{eqtableau}{1}
\omega &= \frac{-i}{8\pi G} \Tr[\delta_1 A \wedge \delta _2 \Sigma-
\delta_2 A \wedge \delta _1 \Sigma]
-\frac{i}{8\pi G}d[\delta_{1}\psi\,\delta_{2}
\,({}^{2}\!\epsilon) - \delta_{2}\psi \,\,\delta_{1}\, 
({}^{2}\!\epsilon)] \\
&\quad  + \frac{1}{4\pi} \delta_1 \emA \wedge \delta_2 \dual\emF
    - \delta_2 \emA \wedge \delta_1 \dual\emF
\end{eqtableau}
\end{equation}
Note that the exact differential in the expression of $\omega$ vanishes at
infinity due to the fall-off conditions on $A$.  Hence, on integrating over
$M$, it provides just the surface term at the horizon in the expression
(\ref{Sym}) of the symplectic structure.  The expression of the 3-form
$\omega$ involves 4-dimensional fields.  However, when we integrate it over
$M$ to obtain the symplectic structure, only the pull-backs to $M$ of these
fields contribute.

The main idea behind our calculation can be summarized as follows.  When
equations of motion are satisfied, the 3-form $\omega$ is curl-free. 
Therefore, the integral of $d\omega$ trivially vanishes in the 4-dimensional
region $\man$ bounded by the isolated horizon $\Delta$, a partial Cauchy
surface $M$ and null infinity $\scri^+$.  Hence, provided all fields remain
regular (in a conformal completion in which $i^+$ is a single point), the
integral of $\omega$ on $M$ equals the sum of the integrals over $\Delta$ and
$\scri^+$.  Let us first consider the integral over $\Delta$.  Using the
isolated horizon boundary conditions one can show the sum of the first and
last terms can be expressed as an exact differential which is precisely the
negative of the one appearing in the second term of $\omega$.  Thus, the
integral of the symplectic current over $\Delta$ vanishes.  Hence, the
integral of the symplectic current over $M$ equals that over $\scri^+$.  The
former is just the canonical symplectic structure (\ref{Sym}).  The idea now
is to show that the latter is the radiative symplectic structure at $\scri^+$.

Let us therefore evaluate $\int_{\scri^+} \omega$.  It is immediate from the
falloff condition $A \sim O(\frac{1}{r^2})$ that the exact part (i.e., the
second term in the expression) of $\omega$ does not contribute at $\scri^+$.
Next, using the conformal invariance of Maxwell's equations, it is fairly
straightforward to evaluate the electro-magnetic part (i.e. the third term)
of this integral.  Since $\hat{\emA} = \emA$ satisfies Maxwell's equations
on $(\hat{\man}, \hat{g}_{ab})$, the Maxwell potentials $\emA$ are
well-behaved at $\scri^+$.  Therefore, we impose the gauge condition
(\ref{emACond}) at $\scri^+$ and evaluate the electro-magnetic contribution to
the integral of $\omega$ over $\scri^+$.  It equals precisely the
electro-magnetic part of the symplectic structure (\ref{SymScri}) at
$\scri^+$.

Thus, the non-trivial part of the calculation lies in integrating the first
term in the symplectic current over $\scri^+$.  We will now sketch the main
steps.

The gravitational symplectic structure (\ref{Sym}) is expressed in terms of
the fields $\Sigma$ and $A$.  To compare it with the symplectic structure at
$\scri^+$, we first need to re-express it in terms of the metric
$\hat{g}_{ab}$ and its variations.  Since we are assuming the equations of
motion, and in particular Gauss' law $D\Sigma =0$, the connection $A$ can be
expressed in terms of the soldering form $\sigma$ as
\begin{equation}
\tensor<A_a^AB>=-\frac{1}{2}\tensor<\sigma^bAA'> \nabla_a
\tensor <\sigma_b^B_A'>.
\end{equation}
An arbitrary variation to the soldering form $\sigma$ can be written
as
\begin{equation}
\delta\tensor<\sigma_a^AA'>=\frac{1}{2}(\delta
g_{ab}\tensor<\sigma^bAA'> + \mu_{ab}\tensor<\sigma^bAA'>)
\end{equation}
where $\delta g_{ab}$ is symmetric and $\mu_{ab}$ is antisymmetric.  It is
easy to check that the above variation in $\sigma$ induces a variation $\delta
g_{ab}$ in the metric.  Also, by performing an internal gauge transformation
on $\delta \sigma$ {\it{without}} changing the background field $\sigma$,
$\mu_{ab}$ can be set equal to zero.  This gauge transformation leaves the
variation of the metric $\delta g_{ab}$ unchanged.  Hence, from now on,
without loss of generality, we assume we are in a gauge in which $\mu_{ab}$
has been set to zero.  This choice of internal gauge will simplify our
calculations considerably but is \textit{not} essential since the symplectic
structure is gauge invariant.  We can now express $\delta \sigma$ in terms of
the conformally rescaled soldering form $\hat{\sigma}_a=\Omega\sigma_a$ as
\begin{equation}
\delta\tensor<\sigma_a^AA'>
=\frac{1}{2\Omega}\delta {g}_{ab}\tensor<\hat{\sigma}^bAA'>
\end{equation}

Now, one would naively expect $\Omega^2\, \delta g_{ab}$ to be finite at
$\scri^+$.  However, in the context of vacuum general relativity, Geroch
and Xanthopoulos \cite{gx} have shown that perturbations with
$C^\infty$ initial data of compact support have a better behavior: in a
suitable gauge, $\Omega^2\, \delta g_{ab}$ in fact vanishes at
$\scri^+$. Furthermore,
\begin{equation}\label{GX}
  h_{ab}:= \Omega \, \delta g_{ab}
\end{equation}
is well defined and $C^\infty$ at $\scri^+$ and satisfies the following
conditions:
\begin{equation}\label{GX2}
  \Omega^{-1}\hat{n}^a h_{ab} \quad\mbox{and}\quad
  \Omega^{-2}\hat{n}^a \hat{n}^b h_{ab} \quad\mbox{are $C^\infty$,
and}\quad
  \hat{g}^{ab}\Lie_{\hat{n}}h_{ab}=0
\end{equation}
at $\scri^+$.  Finally, the trace-free part of the field $h_{ab}$ at $\scri^+$
is precisely the field $\delta\gamma_{ab}$ representing the change in the
equivalence class of connections at $\scri^+$, i.e. the tangent vector to the
phase space of radiative modes induced by $\delta g_{ab}$ \cite{am2}
\begin{equation}\label{gammah}
  \delta\gamma_{ab} = h_{ab} -\frac{1}{2} \hat{q}^{mn}h_{mn}\hat{q}_{ab}.
\end{equation}
We will {\it assume} the tangent vectors $\delta$ under consideration
have this asymptotic behavior.

With this structure at hand, a straightforward but lengthy
calculation enables one to express the variations of $A$ and $\Sigma$
in terms of fields which are smooth at $\scri^+$:
\begin{equation}
\begin{eqtableau}{1}
\delta \tensor<A_c^AB> &=\frac{1}{4}
\tensor<\hat{\Sigma}^efAB>
[\Omega (\hat{\nabla}_e h_{cf})
+\hat{g}_{cf} h_{ek}\hat{n}^k] \\
\delta \tensor<\Sigma_ab^AB> &=   \Omega^{-1}
\tensor<\hat{\Sigma}_[a^dAB>h_{b]d}
\end{eqtableau}
\end{equation}
Using the identity
\begin{equation}
\tensor<\Sigma_ab^AB><\Sigma_cdAB>
=2(g_{ac}g_{bd}-g_{ad}g_{bc})-2i\epsilon_{abcd}
\end{equation}
and simple consequences of the Geroch-Xanthopoulos asymptotic behavior
(\ref{GX2}), one can now express the symplectic current in terms of
the fields $h_{1ab}$ and $h_{2ab}$. Assuming that at least one of the two
perturbations, $h_{1ab}$ and $h_{2ab}$ vanishes at $\spi$ and $i^+$, we can
therefore write the gravitational part of the canonical symplectic
structure as
\begin{equation}\label{SymSc}
\Omega(\delta_1,\delta_2) =\frac{1}{32\pi G} \int_{\scri^+}
\tensor<(h_1ab><\hat{n}^c>
<\hat{\nabla}_c><h_2cd> -
\tensor<h_2ab><\hat{n}^c><\hat{\nabla}_c><h_1cd>)
\hat{g}^{ac} \hat{g}^{bd} \,\,
\tensor^\scri<\hat{\epsilon}> \, ,
\end{equation}
where the volume form $\tensor^3<\hat{\epsilon}>$ on $\scri^+$ is given
by (\ref{VolScri}).

To bring this expression to the same form as appears in (\ref{SymScri}) it is
necessary to replace the fields $h_{ab}$ with their trace-free parts
$\delta\gamma_{ab}$ in the first integral.  This will not introduce any
additional terms because of the properties (\ref{GX2}) of $h_{ab}$.  Also,
since $\scri^+$ is divergence-free, we can replace $\hat{n}^c\hat{\nabla}_c
h_{ab}$ with $\Lie_{\hat{n}}h_{ab}$.

In summary, we have shown that, when the equations of motion hold,
both the gravitational and electro-magnetic parts of the symplectic
structure can be rewritten in terms of fields living at $\scri^+$.
Combining these results, it follows that
\begin{equation} \label{symeq}
  \Omega(\delta_1,\delta_2)=\Omega^\Rad\,(\delta^\Rad_1,\delta^\Rad_2),
\end{equation}
provided the background and the tangent vectors have certain
asymptotic properties.

To conclude, let us collect the assumptions on the behavior of various fields
that were necessary to arrive at (\ref{symeq}).  The background solution is
assumed to be asymptotically flat at spatial and future null infinity and
asymptotically Schwarzschild at future time-like infinity.  In a conformal
frame in which $\scri^+$ is divergence-free, the linearized fields $h_{ab}$
are assumed to satisfy the Geroch-Xanthopoulos conditions (\ref{GX}),
(\ref{GX2}) and the Maxwell potential $\delta{\emA}$ is assumed to satisfy
(\ref{emACond}) at $\scri^+$.  Next, at least one of $h_{1ab}$ and $h_{2ab}$
has to vanish at $i^+$ and at least one of them has to vanish at $\spi$.  This
last assumption can easily be met in the actual application of (\ref{symeq})
in the main text (Section \ref{s7.3}).  There, $\delta_2$ is the Hamiltonian
vector field associated with a BMS time translation $\hat{n}^a$; $\delta_2 =
(\Lie_{\hat{n}}\gamma_{ab},\, \Lie_{\hat{n}}\emA)$.  Now, up to numerical
factors, the total energy radiated across $\scri^+$ in the (background)
space-time is given by the integral of squares of these two fields.  Hence, it
is physically reasonable to restrict oneself to space-times in which the two
fields go to zero as one approaches $i^+$ and $\spi$ along $\scri^+$.  In this
case, $h_{2ab}$ will automatically satisfy the last requirement.  Finally, for
the main result (\ref{HeqERad}) of Section \ref{s7.3} to hold, an additional
condition must be satisfied: the linearized fields $h_{ab}$ and $\delta\emA$
satisfying (\ref{GX}), (\ref{GX2}) and (\ref{emACond}) should span the tangent
space at each point of the sector of phase space considered.

While these assumptions seem plausible, we do not know of general results
which will ensure that a `sufficient number' of such background solutions
exist or that they will admit a `sufficient number' of linearized fields
satisfying our conditions.  Indeed, at this stage, one does not even have a
conclusive proof of existence of a `sufficient number' of radiating
solutions which have smooth and complete $\scri^+$.

\end{document}